\documentclass[10pt]{asme2ej}

\usepackage{epsfig} 
\usepackage{subfig}
\usepackage{array} 
\usepackage{multirow}
\usepackage{color}
\newcolumntype{L}[1]{>{\raggedright\let\newline\\\arraybackslash\hspace{0pt}}m{#1}}
\newcolumntype{C}[1]{>{\centering\let\newline\\\arraybackslash\hspace{0pt}}m{#1}}
\newcolumntype{R}[1]{>{\raggedleft\let\newline\\\arraybackslash\hspace{0pt}}m{#1}}
\newcolumntype{N}{@{}m{0pt}@{}}
\title{Verification and Validation: the Path to Predictive Scale-Resolving Simulations of Turbulence}

\author{Filipe S. Pereira
    \affiliation{
	Postdoc Researcher Associate\\
	T - Theoretical Division\\
	Los Alamos National Laboratory\\
	Los Alamos, New Mexico 87544\\
    Email: fspereira@lanl.gov
    }	
}

\author{Fernando F. Grinstein 
    \affiliation{
        Scientist\\
	X - Computational Physics Division\\
	Los Alamos National Laboratory\\
	Los Alamos, New Mexico 87544\\
    Email: fgrinstein@lanl.gov
    }
}

\author{Daniel M. Israel
    \affiliation{Scientist\\
	X - Computational Physics Division\\
	Los Alamos National Laboratory\\
	Los Alamos, New Mexico 87544\\
    Email: dmi1@lanl.gov
    }
}

\author{Luis E\c{c}a
    \affiliation{Associate Professor\\
	Department of Mechanical Engineering\\
	Instituto Superior T\'ecnico\\
	Lisbon, Portugal\\
    Email: luis.eca@ist.utl.pt
    }
}

\begin{document}

\maketitle    

\begin{abstract}
{\it 
This work investigates the importance of verification and validation (V\&V) to achieve predictive scale-resolving simulations (SRS) of turbulence, i.e., computations capable of resolving a fraction of the turbulent flow scales. Toward this end, we propose a novel but simple V\&V strategy based on grid and physical resolution refinement studies that can be used even when the exact initial flow conditions are unknown, or reference data are unavailable. This is particularly relevant for transient and transitional flow problems, as well as for the improvement of turbulence models. We start by presenting a literature survey of results obtained with distinct SRS models for flows past circular cylinders. It confirms the importance of V\&V by illustrating a large variability of results, which is independent of the selected mathematical model and Reynolds number. The proposed V\&V strategy is then used on three representative problems of practical interest. The results illustrate that it is possible to conduct reliable verification and validation exercises with SRS models, and evidence the importance of V\&V to predictive SRS of turbulence. Most notably, the data also confirm the advantages and potential of the proposed V\&V strategy: separate assessment of numerical and modeling errors, enhanced flow physics analysis, identification of key flow phenomena, and ability to operate when the exact flow conditions are unknown or reference data are unavailable.
}
\end{abstract} 

%
%
%
\section{Introduction}
\label{sec:1}

Scale-resolving simulation (SRS) models are nowadays widely applied in computation of turbulent flow problems of practical interest, e.g., vehicle aerodynamics and hydrodynamics \cite{FUREBY_OE_2016,NISHIKAWA_JJSNAOE_2012,KRAJNOVIC_JFE_2005,MINELLI_JFE_2018,SCHAUERHAMER_AIAA_2016,BENSOW_JSR_2019,ALIN_JSR_2010}, offshore engineering \cite{ZHANG_OMAE29_2010,LEFEVRE_OMAE2020_2013,LIY_POF_2017,PEREIRA_JFE_2019,PEREIRA_IJHFF_2019,WANG_JMSA_2020}, propeller design \cite{LU_JMST_2014,KUMAR_JFM_2017,GUILMINEAU_JFE_2018,PEREIRA_AIAA_2019,ASNAGUI_OE_2020,LIEBRAND_JFE_2021}, materials mixing \cite{DIMONTE_POF_2004,THORNBER_POF_2011,BANERJEE_IJHMT_2009,PEREIRA_PRF2_2021,PEREIRA_PRE_2021,GRINSTEIN_CAF_2020}, and combustion \cite{DEBRUYNKOPS_JFE_2000,DE_JEGTP_2009,HASSE_IJER_2015,GS_AIAA46_2016,PUGGELLI_JEGTP_2017,LEUDESDORFF_MTZ_2019}. These formulations are characterized by their ability to resolve a fraction of or all turbulence scales present in a flow. This property unleashes the potential of SRS methods to obtain high-fidelity predictions of complex flows not amenable to modeling with the Reynolds-averaged Navier-Stokes (RANS) equations due to their limitations representing transient phenomena, onset and development of turbulence, instabilities, and coherent structures. Among all SRS formulations, hybrid \cite{SPALART_FORSR_1997, SPALART_TCFD_2006, SHUR_IJHFF_2008, KOK_AIAA42_2004, MENTER_SHRLM_2016, KOK_FTC_2017}, bridging \cite{SPEZIALLI_ICNMFD_1996, FASEL_JFE_2002, MENTER_AIAA41_2003, MENTER_FTC_2010, GIRIMAJI_JAM_2005, SCHIESTEL_TCFD_2005, CHAOUAT_PF_2005}, and large-eddy simulation (LES) \cite{SMAGORINSKY_MWR_1963,BORIS_FDR_1992,GRINSTEIN_BOOK_2010} are engineering practitioners' most usual modeling strategies. 

The further establishment of SRS methods in many areas of science and engineering is inevitably dependent on the ability to obtain predictive computations, i.e., simulations with a quantified and acceptable level of numerical, input, and modeling uncertainty \cite{ASME_BOOK_2009,PEREIRA_ACME_2021}. Yet, the estimation of these sources of computational uncertainty is difficult in SRS and, as such, often neglected. There are four primary reasons for this: 
\begin{itemize}
\item[\textit{i)}] \textbf{Complexity} - SRS models calculate instantaneous variables which are closely dependent on the physical resolution or range of scales resolved by the model. This hampers the development of accurate and robust SRS turbulence closures by making the comparison and interpretation of some quantities difficult, as it is not possible to match the physical resolution of the reference experiments or simulations.
\item[\textit{ii)}] \textbf{Cost} -  SRS calculations are usually computationally more intensive than RANS owing to the numerical requirements (spatio-temporal grid resolution, simulation time, iterative convergence criterion, etc.) necessary to accurately resolve a fraction of the turbulence field. This makes detailed V\&V exercises time consuming and resource intensive.
\item[\textit{iii)}] \textbf{Dynamic modeling} - the physical resolution of most SRS formulations is defined based on the grid resolution and the simulated flow properties to optimize the use of the available computational resources. However, such modeling option makes the formulation's physical resolution vary upon grid refinement. This turns SRS models grid dependent, precluding the evaluation of modeling and numerical errors separately, a requirement for reliable V\&V exercises. Also, dynamic SRS formulations lead to commutation errors since the filtering operator of the model does not commute with spatial and temporal differentiation \cite{HAMBA_PF_2011}.
\item[\textit{iv)}] \textbf{Numerical setup} - a precise validation exercise requires numerical simulations and reference experiments performed on the same flow conditions, i.e., domain, boundary conditions, material properties, and initial or inflow conditions. For SRS it is not enough to know the mean values of these quantities; initializing the model requires both the time resolved large scale motions and the detailed statistics of the small scales.  However, there are many cases where some of this information is not available from the reference experiments. This is particularly important for transient and transitional flows and may significantly affect proper comparisons between simulations and reference experiments.
\end{itemize}

\begin{figure}[t!]
\centering
\includegraphics[scale=0.25,trim=0 0 0 0,clip]{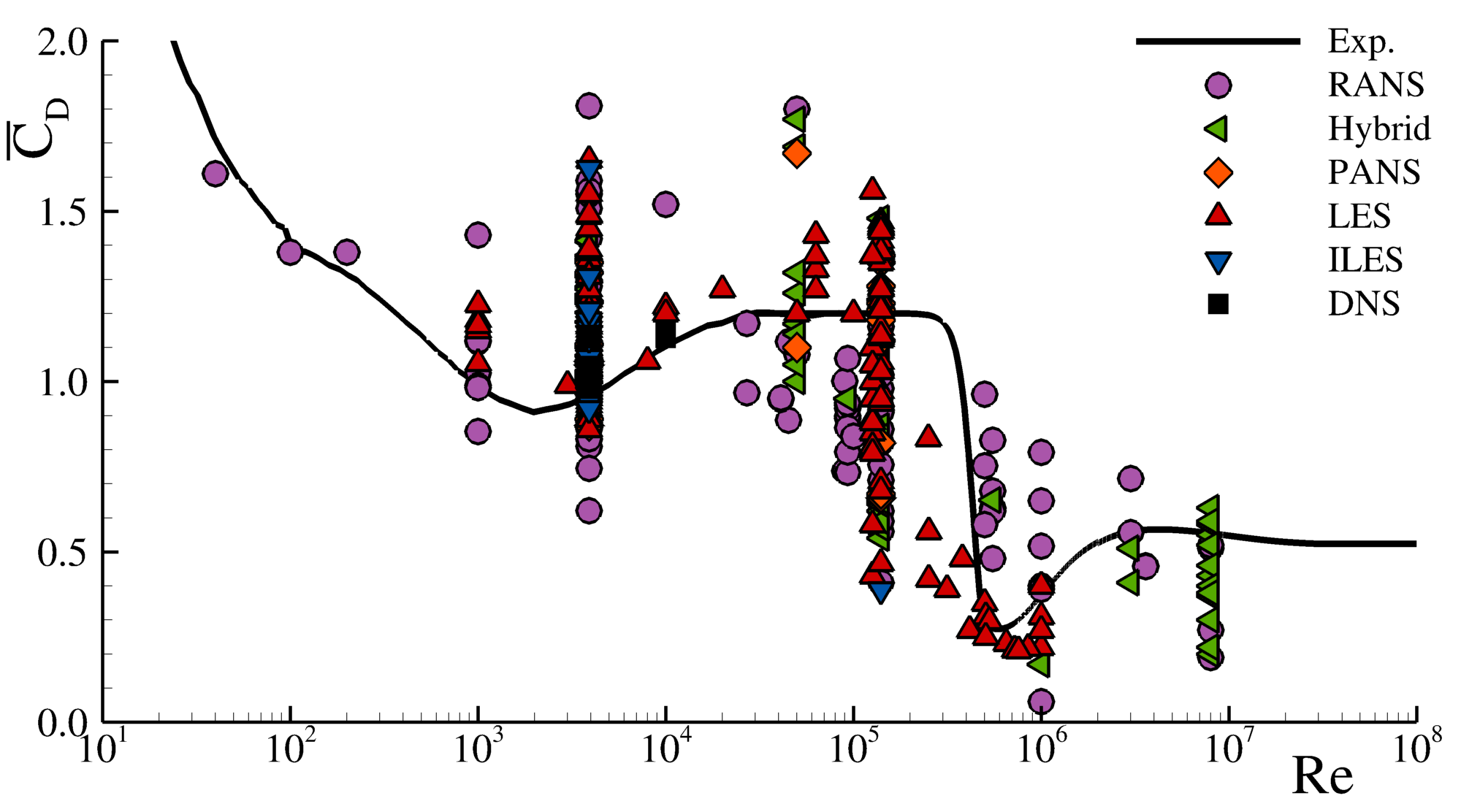}
\caption{Literature survey of numerical results for the time-averaged drag coefficient, $\overline{C}_D$, for the flow around a circular cylinder at different $\mathrm{Re}$ \cite{PEREIRA_PHD_2018,PEREIRA_ACME_2021}.}
\label{fig:1_1}
\end{figure}

To illustrate the role of V\&V exercises \cite{ROACHE_BOOK_1998,ASME_BOOK_2009,ASME_BOOK_2016,OBERKAMPF_BOOK_2010} to numerical prediction and turbulence modeling, figure \ref{fig:1_1} depicts the outcome of a literature survey \cite{PEREIRA_PHD_2018,PEREIRA_ACME_2021} that compiles results for the time-averaged drag coefficient obtained for the flow around a circular cylinder at different Reynolds numbers, Re. One set of experimental measurements is included for comparison \cite{ESDU_1986}, and the numerical results are colored by category of model: direct numerical simulation (DNS), implicit large-eddy simulation (ILES), LES, the bridging partially-averaged Navier-Stokes (PANS) equations, hybrid, and RANS. Note that except for RANS, all these formulations are SRS models. The numerical results exhibit an enourmous range of values, which does not appear to improve for any specific mathematical model.  Nor does it converge with physical resolution, from RANS where turbulence is completely modeled, to LES/ILES where most turbulent scales are resolved. Considering the case $\mathrm{Re}=3900$, the collected $\overline{C}_D$ results  vary from $0.6$ to $1.8$, representing a maximum variation of $191\%$ (taking $\overline{C}_D=0.6$ as the reference). Overall, figure \ref{fig:1_1} emphasizes the need for V\&V exercises to further establish SRS methods in engineering, and generate confidence in their numerical results when reference solutions are unavailable.

This study investigates the importance of V\&V exercises to achieve predictive SRS of turbulence. This is a necessary step to diminish the variability of SRS results in figure \ref{fig:1_1}, and gain confidence on such formulations. Toward this end, we propose a novel but simple V\&V approach based on grid and physical resolution refinement studies that can be used even when the initial flow conditions are not sufficiently characterized, or reference results are unavailable. This is crucial for transient and transitional flow simulations, and for the improvement of the accuracy of turbulence models. The proposed method is utilized to ascertain the simulations' fidelity of three representative test-cases involving transition to turbulence: the flow around a circular cylinder (CC) at $\mathrm{Re}=3.9\times 10^3$, the Taylor-Green vortex (TGV) at $\mathrm{Re}=3000$ \cite{TAYLOR_PRSA_1937}, and the Rayleigh-Taylor (RT) flow \cite{RAYLEIGHT_PLMS_1882,TAYLOR_PRSA_1950} at Atwood number $\mathrm{At}=0.5$. The first problem is a statistically unsteady flow with well-characterized flow conditions and available experimental results and uncertainties, enabling the estimation of validation uncertainties. The second test-case is a transient flow with exact flow conditions and available reference DNS results. Yet, the numerical uncertainty of these DNS results has not been estimated. The third problem is a transient flow without reference experimental results due to the complexity of characterizing the exact experimental RT flow conditions (see literature survey of Pereira et al. \cite{PEREIRA_POF_2021}). These three test-cases constitute the ideal validation space for the new V\&V method. All computations are based on the PANS equations \cite{GIRIMAJI_JAM_2005,PEREIRA_IJHFF_2018,PEREIRA_PRF2_2021} at constant physical resolution to enable the assessment of numerical and modeling errors separately, and prevent commutation errors.

This paper is structured as follows. Section \ref{sec:3} describes the proposed V\&V strategy, and Section \ref{sec:2} presents the details of the selected test-cases and numerical simulations. Afterward, the results are discussed in Section \ref{sec:4}, and the conclusions are summarized in Section \ref{sec:5}. 
%
%
%
\section{V\&V Strategy}
\label{sec:3}

Before describing the proposed V\&V strategy, let us start by introducing the concepts of error and uncertainty. Following the American Society of Mechanical Engineers (ASME) standard for V\&V \cite{ASME_BOOK_2009,ASME_BOOK_2016}, an error can be defined as the difference between the observed, $\phi$, and the exact or truth, $\phi_o$, solution,
\begin{equation}
\label{eq:3_1}
E(\phi) \equiv \phi - \phi_o \; ,
\end{equation}
where $\phi_o$ depends on the class of errors being quantified. Yet, exact solutions are usually unavailable for most engineering flows and so we estimate the uncertainty of $\phi$, $U(\phi)$, that establishes an interval that should contain the exact solution with a given degree of confidence. In the ASME V\&V 20 standard \cite{ASME_BOOK_2009}, uncertainty is defined with a $\pm$ sign to obtain an interval
\begin{equation}
\label{eq:3.2}
\phi-U(\phi) \leq \phi_o \leq \phi+U(\phi) \; .
\end{equation}
Therefore, unlike an error, uncertainty is always a positive quantity.

The uncertainty is evidently related to the error of a simulation, $E_t(\phi)$, that includes three components:
\begin{equation}
\label{eq:3_3}
E_t(\phi) \equiv E_i(\phi) + E_m(\phi) + E_n(\phi) \; .
\end{equation}
Here, $E_i(\phi)$ is the input or parameter error caused by inexact values of boundary conditions, material properties or initial conditions, $E_m (\phi)$ is the modeling error defined as the difference between the exact solution of the mathematical model and physical truth, and $E_n(\phi)$ is the numerical error which can be divided into a discretization, iterative, round-off, and statistical (unsteady flows like the CC case) component  \cite{ROACHE_BOOK_1998,ASME_BOOK_2009,OBERKAMPF_BOOK_2010,ASME_BOOK_2016}.

The goal of a validation assessment is to estimate $E_m(\phi)$. However, as illustrated by equation \ref{eq:3_3}, the estimation of $E_m(\phi)$ depends on input, $E_i(\phi)$, and numerical, $E_n(\phi)$, errors. Furthermore, the true physical value is obtained from an experiment that also includes an experimental error  $E_e(\phi)$. Therefore, the estimation of the modeling error $E_m(\phi)$ is performed with an interval centered at the difference between the simulation $\phi$ and the experimental data $\phi_e$, (comparison error, $E_c(\phi)=\phi-\phi_e$) and a width defined by the so-called validation uncertainty, $U_{v}(\phi)$ \cite{ASME_BOOK_2009},
\begin{equation}
\label{eq:3_4}
E_c(\phi) - U_v(\phi) \leq  E_m(\phi) \leq  E_c(\phi) + U_v(\phi) \; .
\end{equation}
$U_{v}(\phi)$ depends on the input, $U_i(\phi)$, numerical, $U_n(\phi)$, and experimental, $U_e(\phi)$, uncertainties. For cases where these three uncertainties are independent, we obtain
\begin{equation}
\label{eq:3_5}
U_v(\phi)^2 = U_i(\phi)^2 +  U_n(\phi)^2 + U_e(\phi)^2 \; .
\end{equation}

The estimation of $U_{v}(\phi)$ can be quite challenging for problems in which the initial flow conditions are not well characterized (large $U_{i}(\phi)$). Also, the calculation of $U_{v}(\phi)$ is not possible in problems where reference experiments are unavailable. This is often the case for such problems as transitional flow, high-speed combustion, climate prediction, and full-scale problems. In addition to these difficulties, the governing equations of most SRS models directly depend on the grid resolution. This feature prevents the separate and reliable quantification of numerical and modeling errors ($E_n(\phi)$ and $E_m(\phi)$) until all flow scales are resolved in a DNS.

Therefore, the numerical prediction of complex flows with SRS models raises the following question: how can one conduct reliable V\&V exercises to evaluate the accuracy of the computations? We propose the following strategy. Let us start by considering an SRS formulation whose governing equations do not depend on the grid resolution. This requires a model that uses an explicit type of filtering operator. Hence, one can evaluate modeling and numerical errors separately, and the conditions necessary to derive SRS models and assure scale-invariance \cite{GERMANO_JFM_1992} are satisfied. It also prevents commutation errors \cite{HAMBA_PF_2011}. In this work, we use the PANS equations \cite{GIRIMAJI_JAM_2005} supplemented with turbulence closures for incompressible single fluid \cite{PEREIRA_THMT15_2015,PEREIRA_IJHFF_2018} and variable-density flow \cite{PEREIRA_PRF2_2021}. We emphasize that we could have selected other SRS formulations as long as they use an explicit closure to represent the unresolved flow scales, that can operate at constant physical resolution. This might require model modifications. For example, traditional SRS models utilizing a local characteristic grid size $\Delta_\mathrm{SRS}$ to set the physical resolution (e.g., DES, VLES, XLES, LES, etc.) can be modified to either use an uniform $\Delta_\mathrm{SRS}$ or the $\Delta_\mathrm{SRS}$ distribution of the coarsest grid used in the study. This strategy removes the governing equations dependence from the grid cell size, is simple to implement, and enables decoupling numerical from modeling errors. Alternatively, one can use explicit filtering. Yet, the first approach is better suited for practical simulations of turbulence.

\begin{figure*}[t]
\centering
\subfloat[$n=1$]{\label{fig:3.1_0a}
\includegraphics[scale=0.073,trim=0 0 0 0,clip]{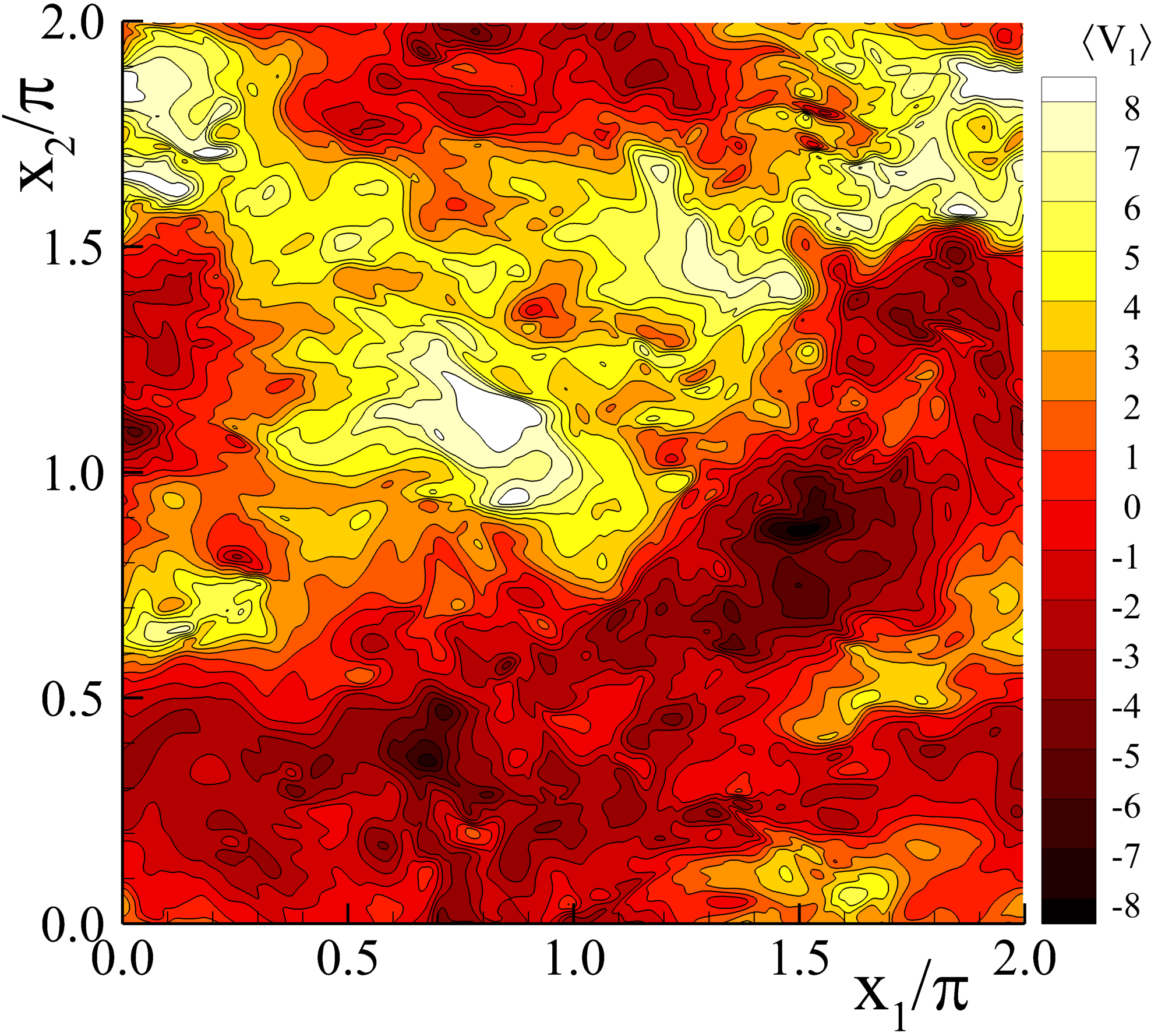}}
~
\subfloat[$n=17$]{\label{fig:3.1_0b}
\includegraphics[scale=0.073,trim=0 0 0 0,clip]{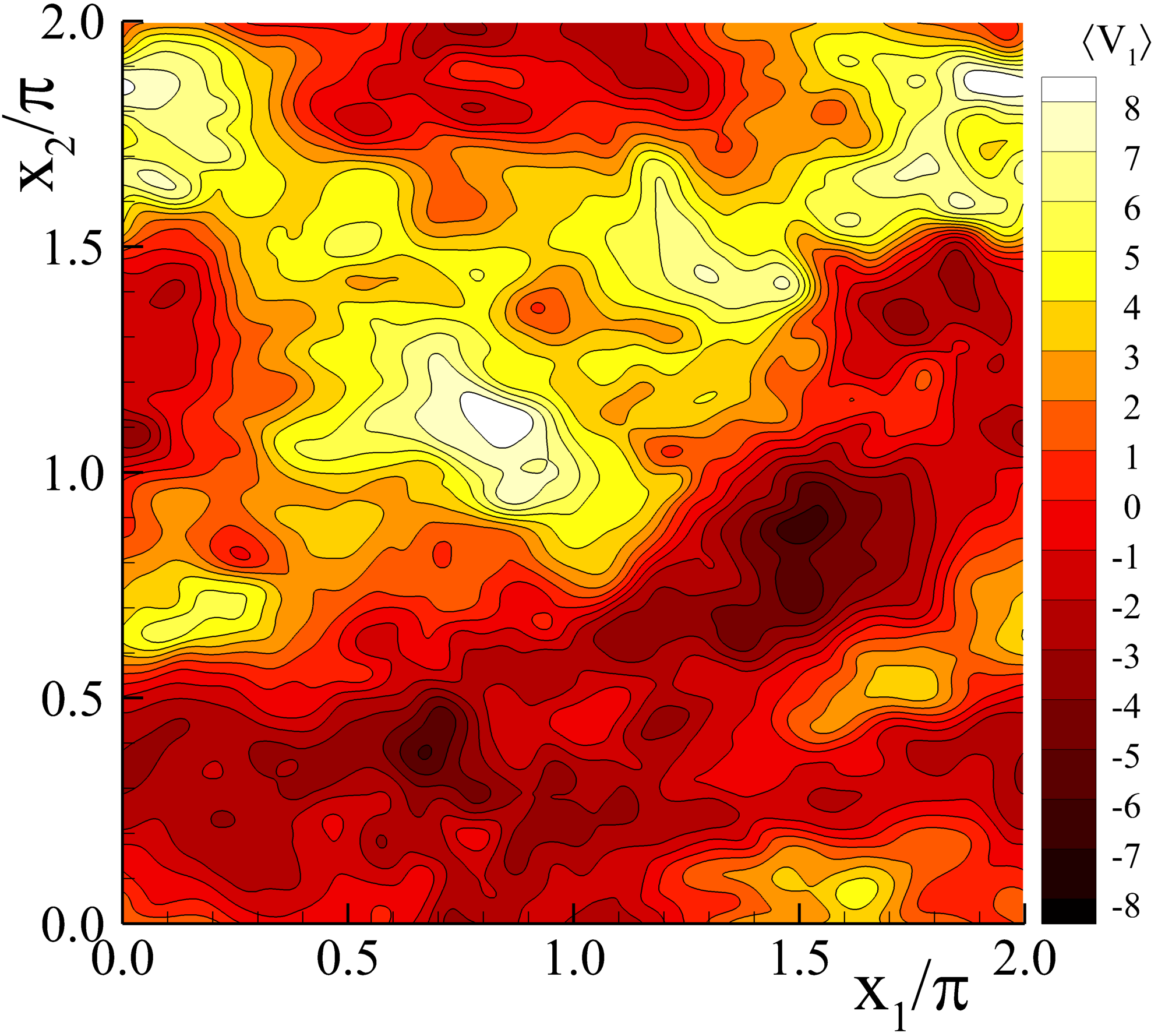}}
~
\subfloat[$n=33$]{\label{fig:3.1_0c}
\includegraphics[scale=0.073,trim=0 0 0 0,clip]{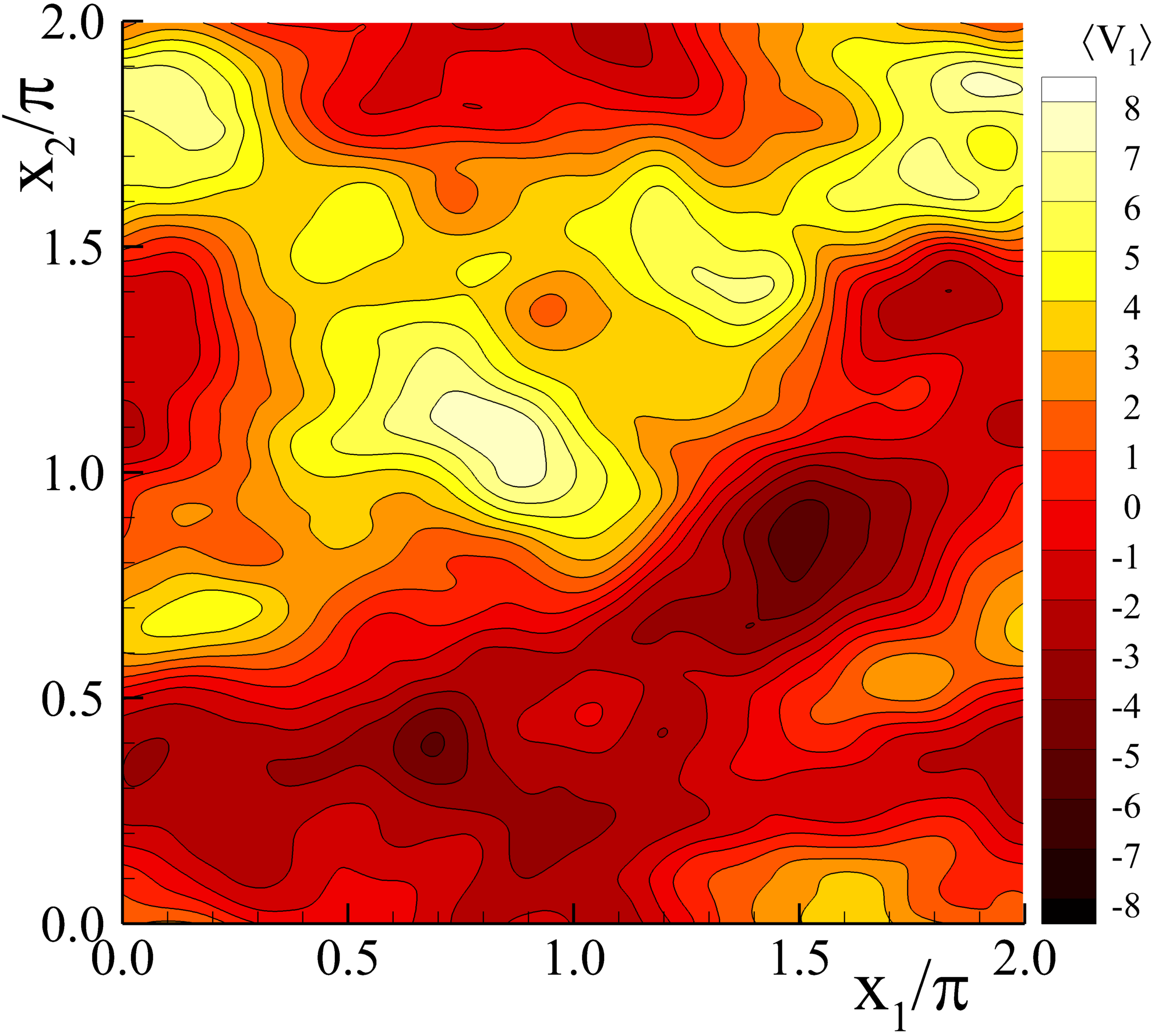}}
\\
\subfloat[$n=69$]{\label{fig:3.1_0d}
\includegraphics[scale=0.073,trim=0 0 0 0,clip]{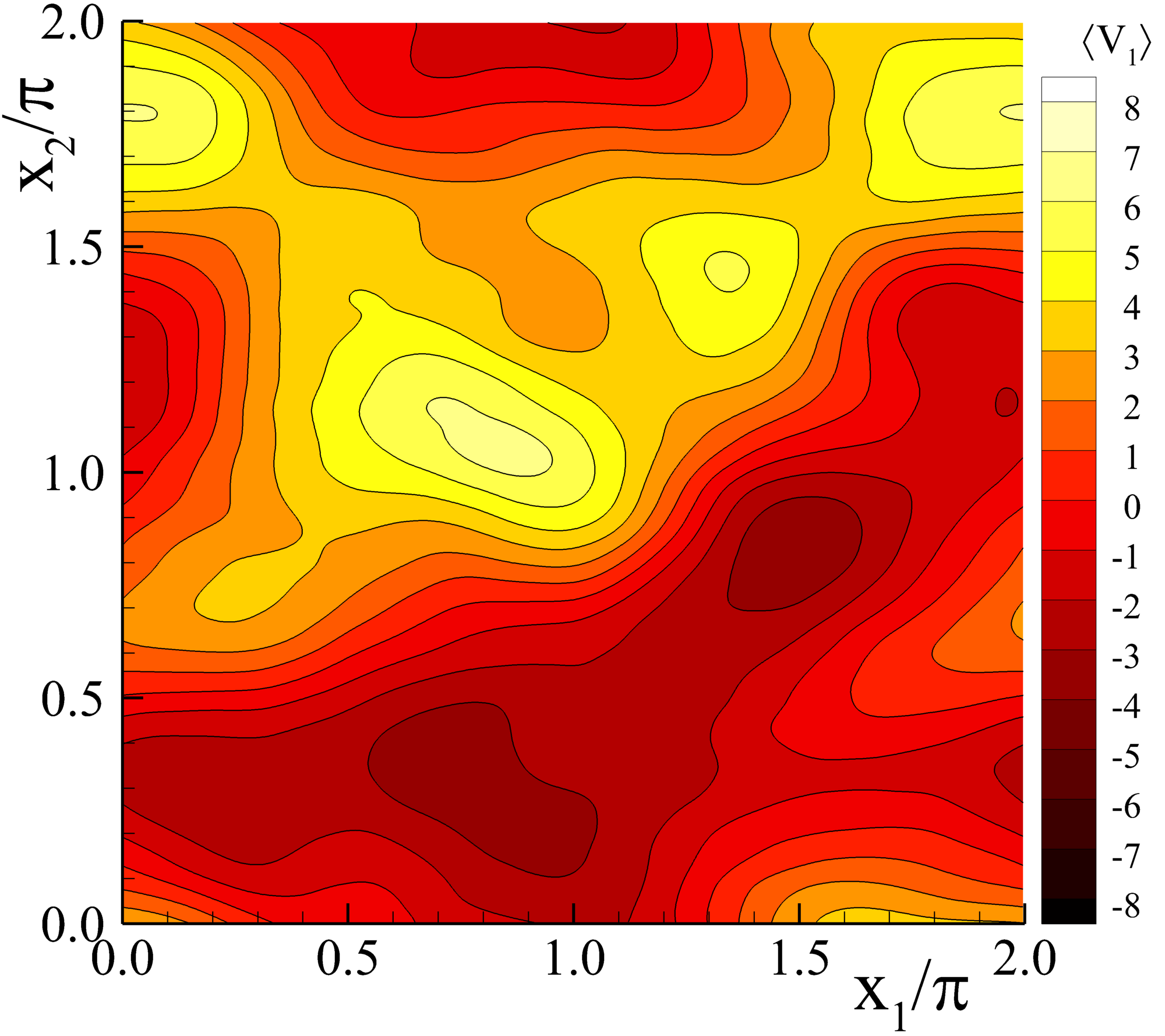}}
~
\subfloat[$n=199$]{\label{fig:3.1_0e}
\includegraphics[scale=0.073,trim=0 0 0 0,clip]{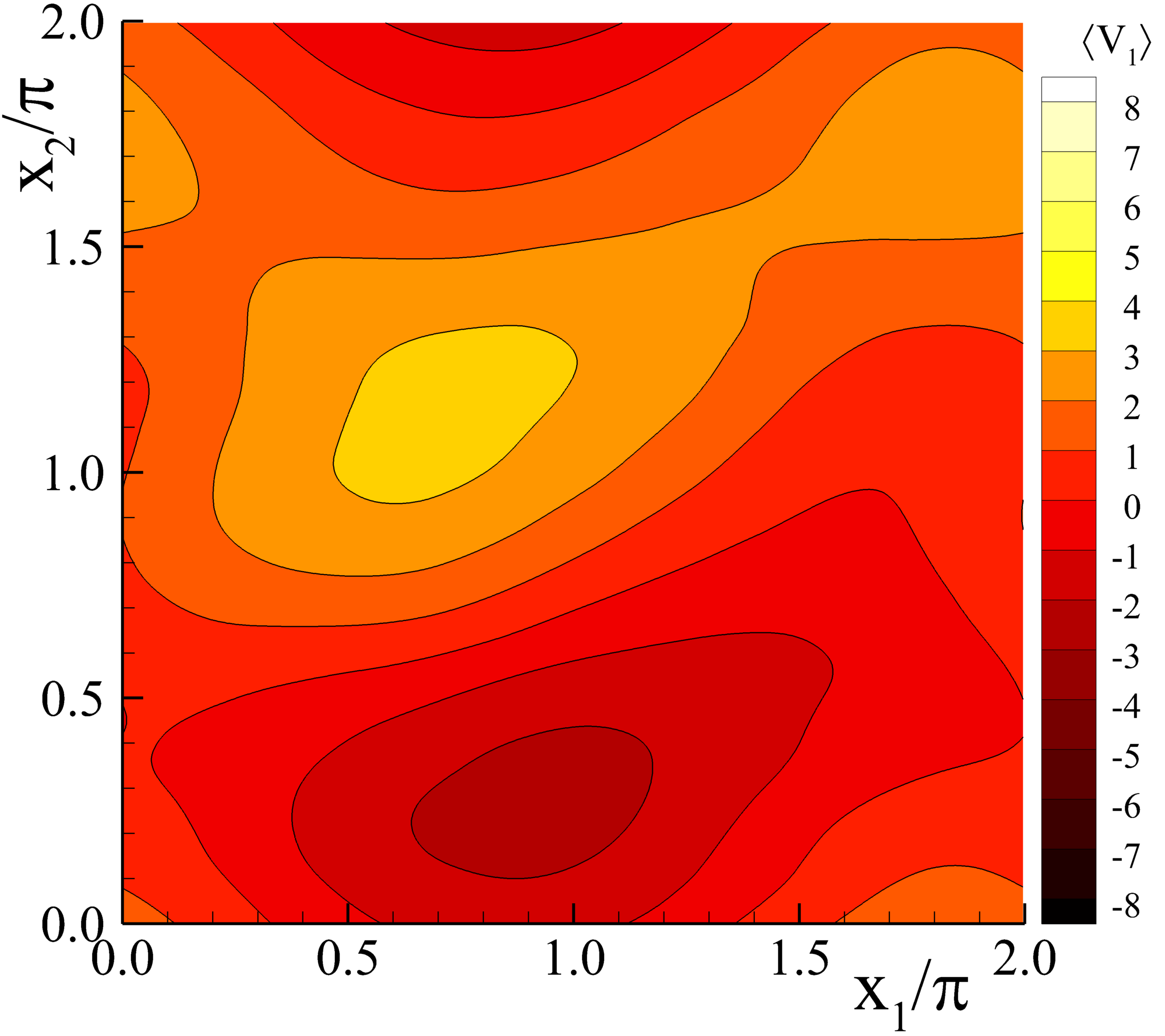}}
~
\subfloat[$n=349$]{\label{fig:3.1_0f}
\includegraphics[scale=0.073,trim=0 0 0 0,clip]{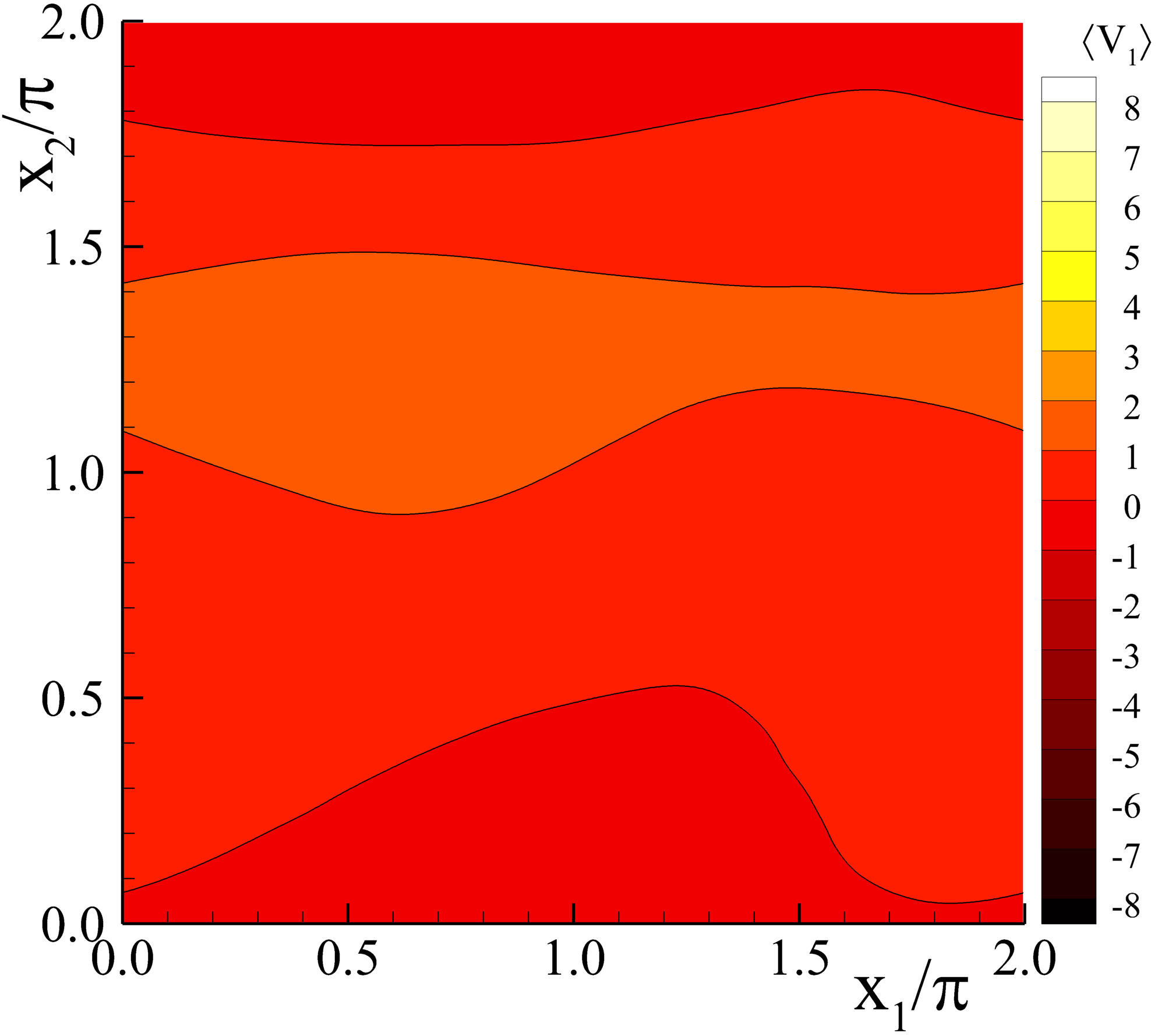}}
\caption{Evolution of $x_1$-component of the forced homogeneous isotropic turbulence flow  \cite{SILVA_JFM_2018} velocity field with the relative filter length size, $n=\Delta/\Delta_\eta$ \cite{PEREIRA_ACME_2021,PEREIRA_PRF2_2021}. $\Delta_\eta$ is the grid size of the DNS simulation.}
\label{fig:3.1_0}
\end{figure*}

In the PANS models of Pereira et al. \cite{PEREIRA_THMT15_2015,PEREIRA_IJHFF_2018,PEREIRA_PRF2_2021}, the physical resolution is set constant, and it is defined by the ratio of the unresolved to the total turbulence kinetic energy,
\begin{equation}
\label{eq:3_6}
f_k= \frac{k_u}{k_t} \; ,
\end{equation}
where $k_u$ is the unresolved or modeled turbulence kinetic energy, and $k_t$ is the total. Note that $f_k=0$ is equivalent to DNS because all turbulence scales are resolved, whereas $f_k=1$ is equivalent to RANS since the turbulence closure models all turbulence scales. For a better physical understanding of the impact of $f_k$ on the predicted flow fields, figure \ref{fig:3.1_0} illustrates how the computed velocity field evolves with the relative filter length size, $n$. $n$ equal to one is a direct numerical simulation ($f_k=0$) where all flow scales are resolved, leading to a highly unsteady velocity field with steep gradients that increase the cost of numerical simulations substantially. In contrast, $n=349$ models most of the turbulence field ($f_k\approx 0.96$), which decreases the cost of the computations.

\begin{figure}[t!]
\centering
\subfloat[$\phi(r_i)$ or $\phi(f_k)$.]{\label{fig:3.1_1a}
\includegraphics[scale=0.147,trim=0 0 0 0,clip]{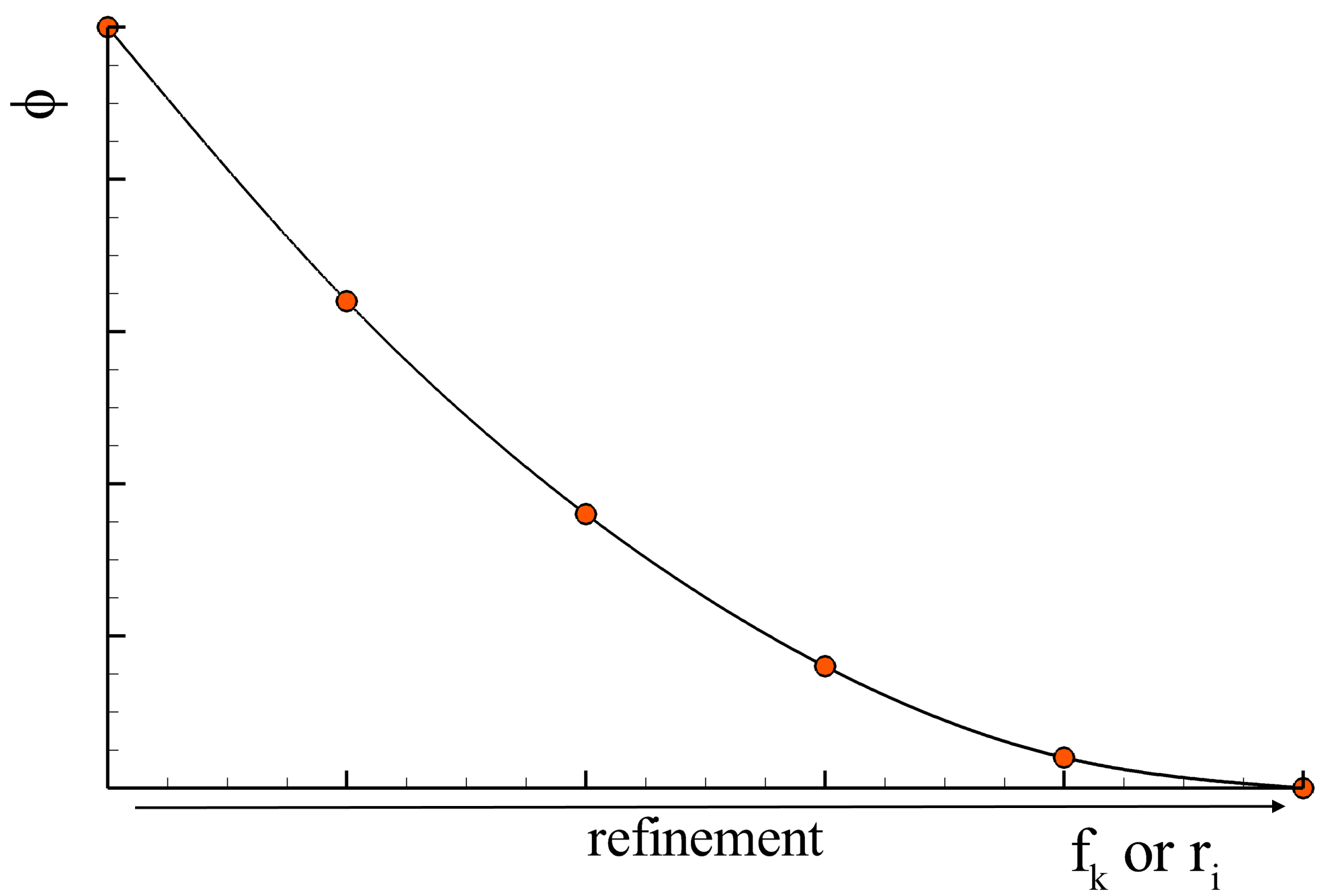}}
~
\subfloat[$\phi(r_i,f_k)$.]{\label{fig:3.1_1b}
\includegraphics[scale=0.160,trim=0 0 0 0,clip]{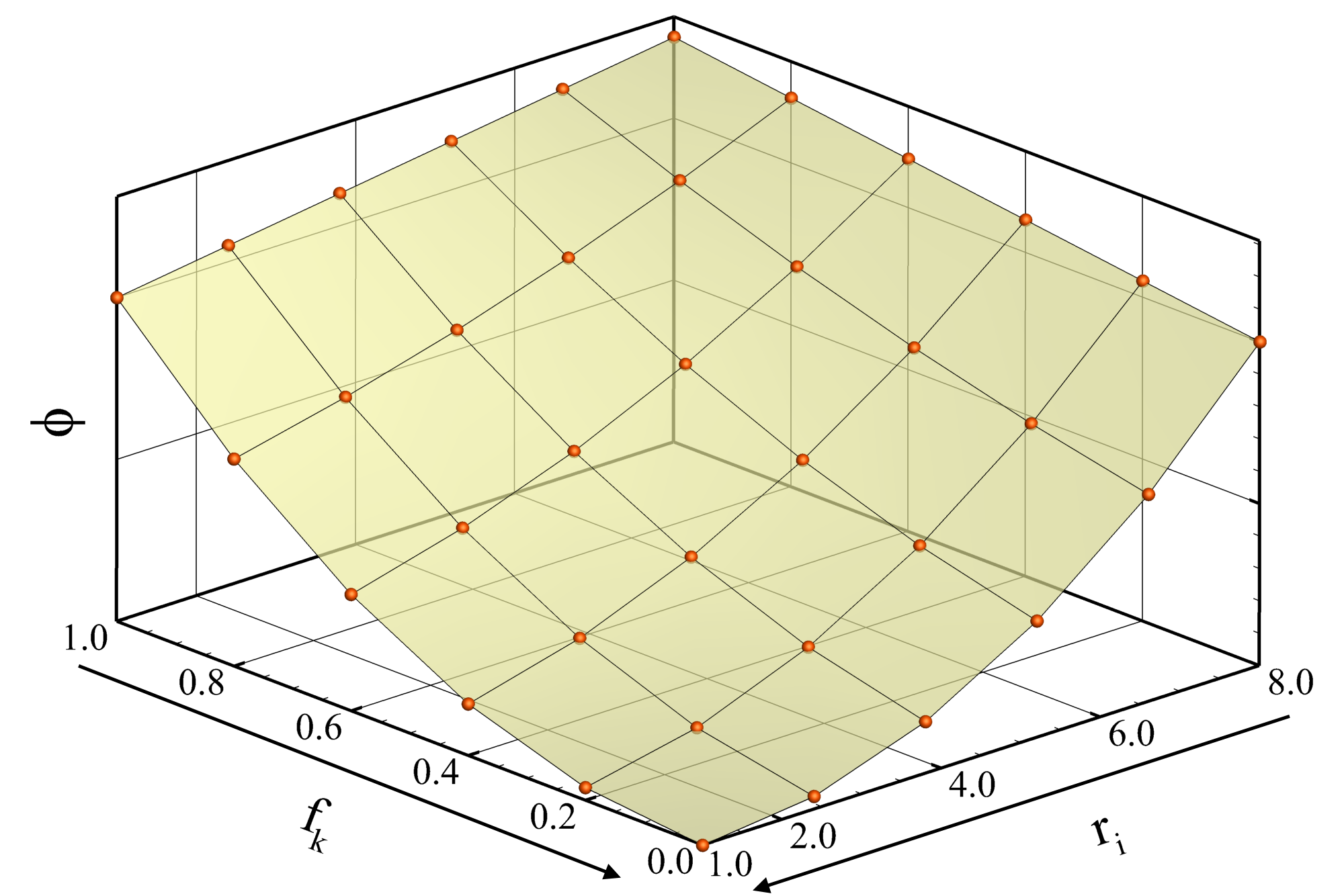}}
\caption{Sketches of generic quantity, $\phi$, evolution upon grid, $r_i$, and physical, $f_k$, resolution refinement.}
\label{fig:3.1_1} 
\end{figure}

This class of SRS formulations provides the background to propose the new V\&V strategy based on grid and physical resolution refinement exercises. It enables the evaluation of the numerical accuracy of the simulations through grid refinement exercises performed at constant $f_k$, and the assessment of  the modeling accuracy of the calculations through physical resolution refinement studies since the solutions converge toward the ``truth'' as $f_k \rightarrow 0$. Yet, the solutions are expected to converge at coarser physical resolutions than the limiting case of $f_k=0$ (DNS for a continuous medium). This is shown below and in numerous studies \cite{PEREIRA_JCP_2018,PEREIRA_IJHFF_2018,PEREIRA_OE_2019,PEREIRA_IJHFF_2019,KAMBLE_POF_2020,FOWLER_POF_2020,PEREIRA_PRF_2021,PEREIRA_POF_2021}, pointing out the method's potential to be applied to complex new problems without experimental solutions. The latter feature is a unique feature of the new V\&V method, which can be crucial to many complex flow problems.

 Figure \ref{fig:3.1_1} illustrates this strategy's typical outcome by depicting the evolution of a generic quantity $\phi$ with the grid refinement ratio,
\begin{equation}
\label{eq:3_7}
r_i=\sqrt[3]{ \frac{N_1}{N_i} }\; ,
\end{equation}
and/or the physical resolution, $f_k$. In relation \ref{eq:3_7}, $N_1$ and $N_i$ are the number of cells of the finest grid available ($i=1$) and a selected grid $i$. The data can be assessed and visualized in three ways: \textit{i)} fixing $f_k$ and plotting $\phi(r_i)$; \textit{ii)} fixing $r_i$ and depicting $\phi(f_k)$; and \textit{iii)} plotting $\phi(f_k,r_i)$. This last approach provides a more comprehensive analysis since it enables the assessment of the numerical and modeling accuracy, their interaction and interdependency, and the interpretation of the predicted flow physics and its dependence on $f_k$ (see Section \ref{sec:4}). Figure \ref{fig:3.1_1} shows that the solutions of $\phi$ converge both with the grid and physical resolutions as $r_i\rightarrow 0$ and $f_k\rightarrow 0$. Note that the smallest $r_i$ shown throughout this work is equal to one, corresponding to the solution obtained on the finest grid tested. $r_i=0$ would require the extrapolation of the solution to an infinitely fine grid (e.g., \cite{ROACHE_BOOK_1998,ECA_JCP_2014}). From the interpretation of figure \ref{fig:3.1_1}, it is possible to infer that the present technique can be utilized  when the experimental flow conditions are poorly characterized or unknown, or in cases in which the reference solution is unavailable. In such cases, one can use the extrapolated solution ($r_i\rightarrow 0$ and $f_k\rightarrow 0$) as reference. It is important to mention that the former procedure could be "generalized" to cases of grid dependent SRS models by considering the cross-derivatives with respect to $r_i$ and $f_k$, $(\partial^2 \phi )/(\partial r_i\partial f_k)$. 

The proposed strategy can evidently be combined with other V\&V techniques \cite{ROACHE_BOOK_1998,CELIK_JFE_2004,DRIKAKIS_AMR_2011,OBERKAMPF_BOOK_2010,PHILLIPS_AIAA20_2011,ECA_JCP_2014,STERN_JFE_2001,CELIK_JFE_2008,RIDER_JCP_2016,XING_JFE_2010,HILLS_JHT_2006,ASME_BOOK_2016,BARMPAROUSIS_IJNMF_2017}. In this work, we estimate the numerical uncertainty of the simulations through the method proposed by E\c{c}a and Hoekstra \cite{ECA_JCP_2014}, and its simplified version utilized in \cite{PEREIRA_IJHFF_2018} (CC flow).
%
%
%
\section{Test-Cases and Numerical Details}
\label{sec:2}
%
%
%
%
\subsection{Flow Problems}
\label{sec:2.1}

We use three representative flow problems to investigate the importance of V\&V to SRS of turbulence, and evaluate the potential of the proposed V\&V strategy. The first problem is a statistically unsteady flow with well-characterized flow conditions and available experimental results and uncertainties, enabling the estimation of validation uncertainties. The second test-case is a transient flow with exact flow conditions and available reference DNS results. Yet, the numerical uncertainty of these DNS results has not been estimated. The third problem is a transient flow without reference experimental results due to the complexity of characterizing the exact experimental initial flow conditions. These three test-cases constitute the ideal validation space for the new V\&V method, and are described below.

\subsubsection{Circular cylinder}
\label{sec:2.1.1}

The flow past a CC at $\mathrm{Re}=3900$ is an archetypal problem commonly used to assess the prediction of transition, turbulence, and separation around bluff-flows in the sub-critical regime \cite{WILLIAMSON_ARFM_1996,ZDRAVKOVICH_BOOK_1997}. This problem is typical of model-scale applications and unmanned vehicles. The flow is statistically unsteady and characterized by a laminar boundary-layer, flow separation leading to a free shear-layer, onset and development of turbulence in the free shear-layer, and finally a turbulent wake. Also, the flow physics is closely dependent on two coherent structures: the Kelvin-Helmholtz in the free shear layer, and the large scale vortex-shedding in the wake. The first is responsible for transition to turbulence in the free shear-layer. The spatial development of the flow in this regime comprises four major steps, proper representation of which determines the accuracy of any numerical simulation: $i)$ onset of the Kelvin-Helmholtz instability in the free shear-layer; $ii)$ spatial development of the Kelvin-Helmholtz rollers; $iii)$ breakdown to high-intensity turbulence; and $iv)$ turbulent free shear-layer roll-up, leading to vortex-shedding. A comprehensive numerical analysis of this problem is given in \cite{PEREIRA_IJHFF_2018,PEREIRA_JCP_2018,PEREIRA_OE_2019}.

In this work, we use the experimental measurements of Parnaudeau et al. \cite{PARNAUDEAU_PF_2008} as reference to evaluate the modeling error. These have been conducted in a wind-tunnel with a square cross section of width equal to $23.3D$ (D is the cylinder's diameter), and an uniform inflow characterized by a turbulence intensity smaller than $I=0.2\%$.  The Reynolds number, $\mathrm{Re} \equiv V_o D/\nu$, is equal to 3900 ($V_o$ is the inflow velocity, and $\nu$ is the kinematic viscosity), and the sampling period used to compute the flow statistics is $\Delta T V_o/D = 2.08\times 10^5$. The quantity of interest extracted from this study is the time-averaged velocity field, $\overline{V}_i$. This has been measured using particle image velocimetry and the reported experimental uncertainties do not exceed $1.0\%$. We also used the experiments of Norberg \cite{NORBERG_BBVIV3_2002,NORBERG_JFS_2003} to assess the predictions of the time-averaged drag coefficient, $\overline{C}_D$, root-mean-square lift coefficient, $C_L'$, and pressure coefficient on the cylinder's surface, $\overline{C}_p(\theta)$. Compared to Parnaudeau et al. \cite{PARNAUDEAU_PF_2008}, these experimental studies are conducted with two slightly different setups: $\mathrm{Re}=4000$, $215.5D\times 80.0D$ cross-section, $I<0.06\%$, and $\Delta T V_o/D = 2.10\times 10^4$ for \cite{NORBERG_BBVIV3_2002}; whereas $\mathrm{Re}=4400$, $314.1D\times 105.0D$ cross-section, and $\Delta T V_o/D = 3.20\times 10^4$ for \cite{NORBERG_JFS_2003}. The reported experimental uncertainties are 0.01 for $\overline{C}_p(\theta)$, $1.0\%$ for $\overline{C}_D$, and $4.0\%$ for $C_L'$.

\begin{figure}[t!]
\centering
\subfloat[CC.]{\label{fig:2.1_1a}
\includegraphics[scale=0.17,trim=0 0 0 0,clip]{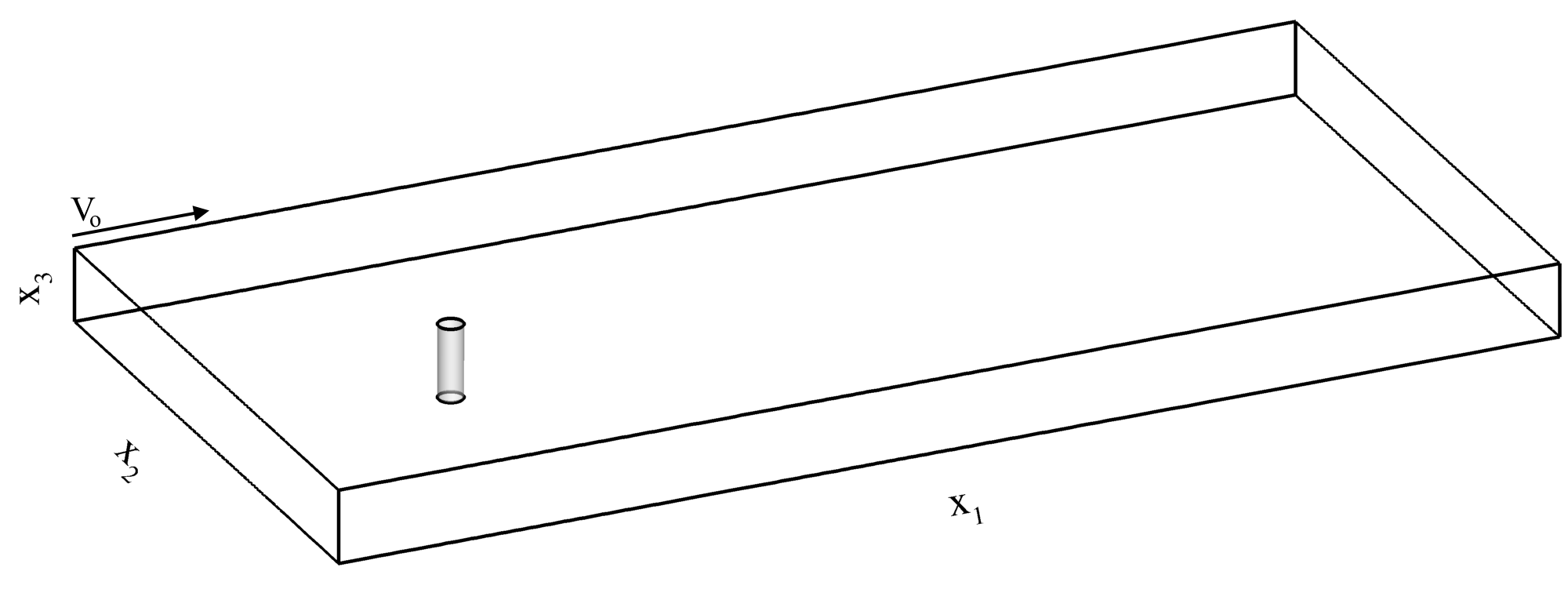}}
\\
\subfloat[TGV.]{\label{fig:2.1_1b}
\includegraphics[scale=0.12,trim=0 0 0 0,clip]{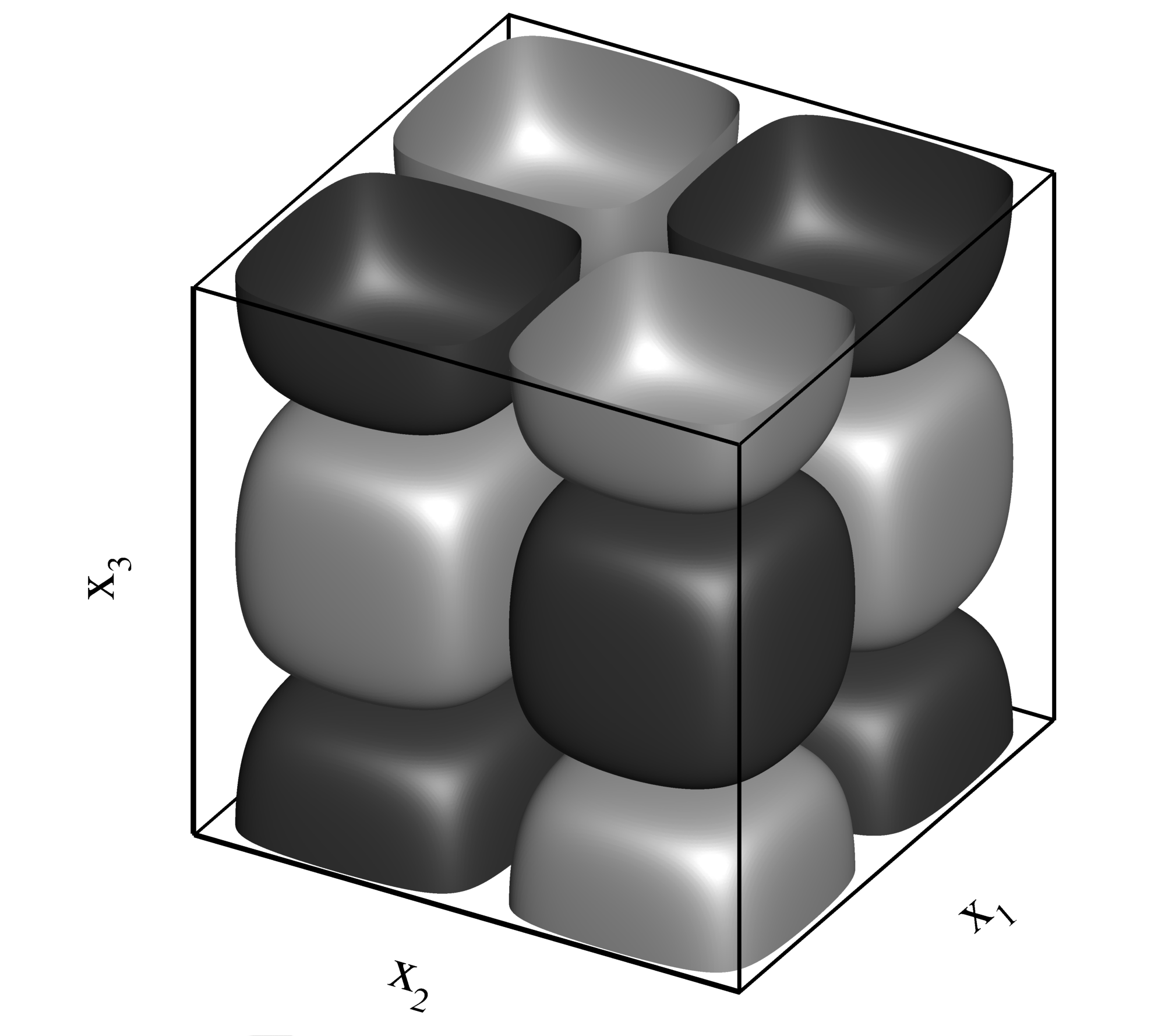}}
~
\subfloat[RT.]{\label{fig:2.1_1c}
\includegraphics[scale=0.07,trim=0 0 0 0,clip]{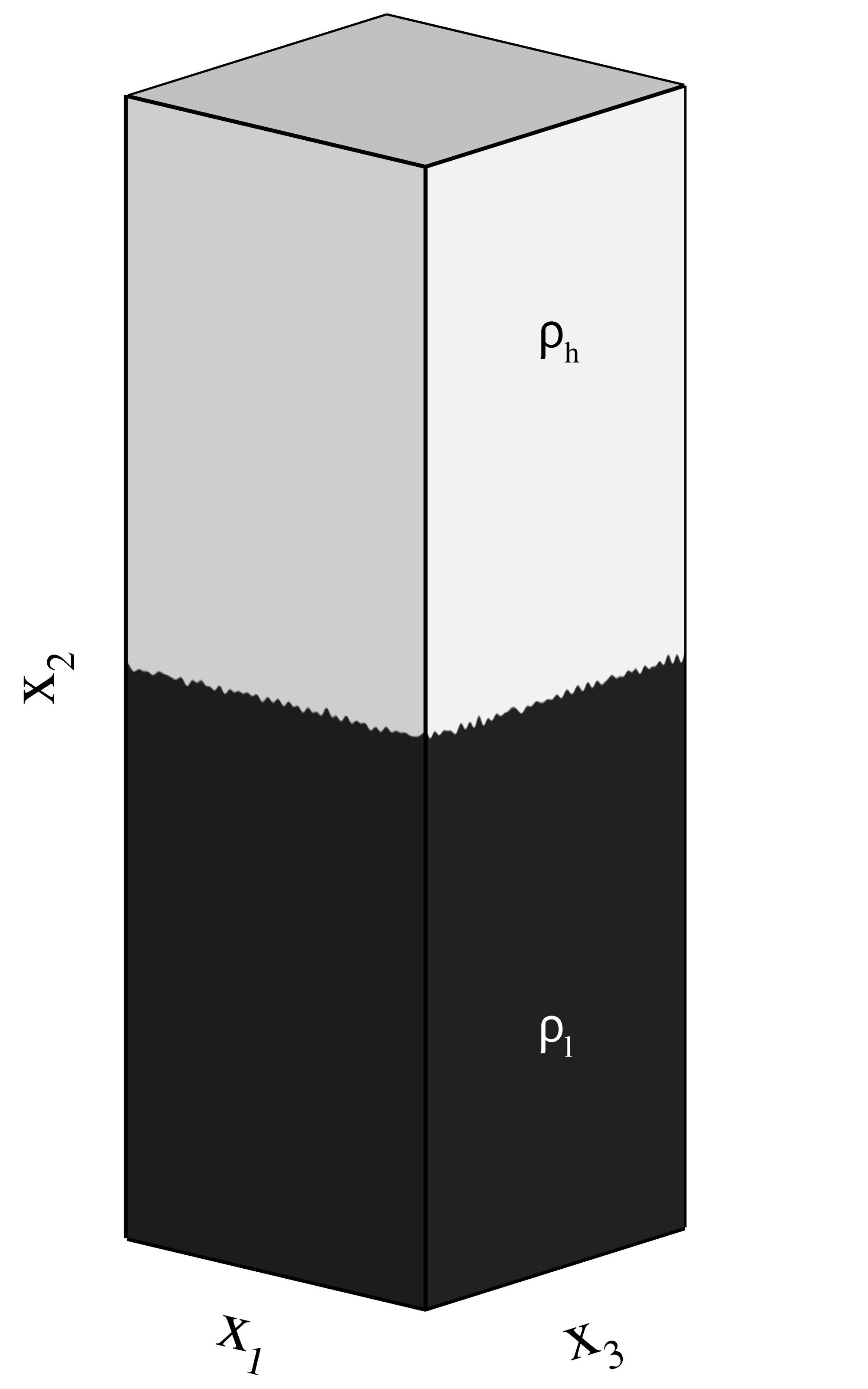}}
\caption{Sketch of the computational domain and initial flow conditions of the CC, TGV, and RT problems.}
\label{fig:2.1_1} 
\end{figure}

The numerical simulations try to replicate the experimental apparatus of Parnaudeau et al. \cite{PARNAUDEAU_PF_2008} closely. The computational domain is shown in figure \ref{fig:2.1_1a}.  The domain is $50D$ long, with the inlet at $10D$ from the inflow, $24D$ wide, and $3D$ thick, with cylinder axis centered. The impact of the cylinder's length ($L=3D$) has been ascertained and guarantees acceptable input uncertainties \cite{PEREIRA_PHD_2018}. At the inlet ($x_1/D=-10$), the prescribed boundary conditions are  constant velocity ($V_o$) aligned with the streamwise direction $x_1$, and the pressure is extrapolated from the interior of the domain. The turbulence quantities are set to result in a turbulence intensity $I=0.2\%$ and a ratio between turbulent and molecular kinematic viscosity equal to $10^{-3}$. At the outlet ($x_1/D=40$), all streamwise derivatives are set equal to zero, whereas no-slip and impermeability conditions are applied on the cylinder's surface. Also, the normal pressure gradient and the turbulence quantities are equal to zero. The exception is the specific dissipation which is given at the center of the nearest-wall cell by $\omega=80\nu d^{-2}$ ($d$ is the nearest-wall distance). The pressure and transverse derivatives of the remaining dependent quantities are equal to zero at $x_2/D=\pm 12$, and symmetry conditions are prescribed at $x_3/D=0$ and $3$.  The simulations are initialized from a $200$ time units ($\Delta T V_o/D$) computation, and run for a minimum of $500$ time units. The flow statistics are calculated using a sampling period $\Delta T V_o/D \ge 350$ \cite{PEREIRA_ACME_2021}.

%
%
%
\subsubsection{Taylor-Green vortex}
\label{sec:2.1.2}

The TGV \cite{TAYLOR_PRSA_1937} is a benchmark transient problem used to investigate the modeling and simulation of transition to turbulence driven by  vortex stretching and reconnection mechanisms \cite{BRACHET_JFM_1983}. The TGV problem starts with a laminar, single Fourier mode, well-defined vortex in a periodic box, illustrated in figure \ref{fig:2.1_1b}. This is given mathematically by \cite{TAYLOR_PRSA_1937,BRACHET_JFM_1983}:
\begin{equation}
\label{2.1.2_1}
V_1(\mathbf{x},t_o)=V_o \sin(x_1)\cos(x_2)\cos(x_3) \; ,
\end{equation}
\begin{equation}
\label{2.1.2_2}
V_2(\mathbf{x},t_o)=-V_o \cos(x_1)\sin(x_2)\cos(x_3) \; ,
\end{equation}
\begin{equation}
\label{2.1.2_3}
V_3(\mathbf{x},t_o)=0 \; ,
\end{equation}
where $V_o$ is the initial velocity magnitude. The corresponding pressure field is obtained from solving the Poisson equation, 
\begin{equation}
\label{2.1.2_4}
P(\mathbf{x},t_o)=  P_o + \frac{\rho_o V_o^2}{16} \left[ 2 +\cos \left( 2x_3 \right)\right] \left[ \cos \left( 2x_1\right) +\cos \left( 2x_2 \right)\right]\; .
\end{equation}
Here, $P_o$ and $\rho_o$ are the pressure and density magnitude at $t=0$. After $t=0$, these structures interact and deform, leading to the formation of vortex sheets. Afterward, the structures get closer and roll-up, causing the onset of turbulence due to a complex vortex reconnection process between pairs of counter-rotating vortices \cite{PEREIRA_PRF_2021}. In the next instants, the remaining vortical structures breakdown, and turbulence further develops. This leads to rapid dissipation of the flow kinetic energy. Since the TGV is an isolated system, the flow kinetic energy decays in time.

The flow configuration of the simulations replicates that used in the DNS studies of Brachet et al. \cite{BRACHET_JFM_1983} and Drikakis et al. \cite{DRIKAKIS_JOT_2007} at $\mathrm{Re}\equiv V_o L_o/\nu=3000$. The simulations are conducted in a cubical computational domain with width $L=2\pi L_o$, where periodic conditions are applied on all boundaries. The initial Mach number is set $\mathrm{Ma}_o=0.28$, and the simulations run until $t=20 V_o/L_o$. The initial thermodynamic and flow properties are the following: vortex velocity $V_o=10^4$cm/s, box size $L_o=1.00$cm, density $\rho_o=1.178\times 10^{-3}$g/cm$^3$, mean pressure $P_o=10^5$Pa, dynamic viscosity $\mu= 3.927 \times 10^{-3}$g/(cm.s), heat capacity ratio $\gamma=1.40$, turbulence kinetic energy $k_o=10^{-7}$cm$^2$/s$^2$, and dissipation length scale $S_o=6.136\times 10^{-3}$cm. The DNS studies of Brachet et al. \cite{BRACHET_JFM_1983} at $\mathrm{Ma}=0$ (incompressible) and Drikakis et al. \cite{DRIKAKIS_JOT_2007} at $\mathrm{Ma}_o=0.28$ are used to evaluate the modeling accuracy of the simulations. These do not report numerical uncertainties. We emphasize that the differences in $\mathrm{Ma}_o$ between our simulations and Brachet et al. \cite{BRACHET_JFM_1983} computations may lead to discrepancies.
%
%
%
\subsubsection{Rayleigh-Taylor}
\label{sec:2.1.3}

The RT \cite{RAYLEIGHT_PLMS_1882,TAYLOR_PRSA_1950} is a benchmark test-case of variable-density flow. This problem is a fundamental test case for flows with material mixing and turbulence generation by baroclinicity. The flow initially consists of two fluids of different densities separated by a perturbed interface. These materials are at rest, with the dense medium ($\rho_h$) located on top of the light material ($\rho_l$) as shown in figure \ref{fig:2.1_1c}. Immediately after $t=0$, the heavy fluid accelerates downward by the action of a body force, whereas the light fluid moves upwards. The interface perturbations create a misalignment between the density gradient and the pressure, which induces the RT instability. This generates structures: penetration of heavy fluid into the light fluid (called bubbles) and, conversely, spikes by penetration of light fluid into the heavy fluid. The boundaries of these structures experience shearing, resulting in a Kelvin-Helmholtz instability, which triggers the onset of turbulence. Afterward, turbulence further develops, enhancing the mixing of the two materials.  The density difference is characterized by the Atwood number, $\mathrm{At}=(\rho_h-\rho_l)/(\rho_h+\rho_l$).

The present flow configuration is the same as that of Pereira et al. \cite{PEREIRA_PRF2_2021,PEREIRA_POF_2021} at At=0.50. The computational domain is a rectangular prism with a squared cross-section of width $L_o=2\pi\, \mathrm{cm}$, and a height of $3L_o$. The mixing-layer height does not exceed $2L_o$ during a simulation time of 20 time units (time normalized by $t^*=\sqrt{L_o/(32g\mathrm{At}  )}$ \cite{LIVESCU_PD_2020}). Periodic conditions are applied on the lateral boundaries, whereas reflective conditions are prescribed on the top and bottom boundaries. The interface is perturbed as in \cite{PEREIRA_POF_2021,PEREIRA_POF_2021} with density fluctuations, which height is defined by,
\begin{equation}
\label{eq:2.1.3_1}
h_p(x_1,x_3) = \sum_{n,m} \cos\left[ 2 \pi \left( n\frac{x_1}{L_o} + r_1\right) \right] \cos\left[ 2 \pi \left( m\frac{x_3}{L_o} + r_3\right) \right] \; .
\end{equation}
The perturbations' wavelengths range between modes 30 and 34 ($30 \leq \sqrt{n^2 + m^2} \leq 34$), and their amplitude is featured by a maximum standard deviation equal to 0.04$L_o$. In equation \ref{eq:2.1.3_1}, the modes $m$ and $n$ are selected to include the most unstable mode of the linearized problem, and $r_1$ and $r_3$ are random numbers between $0$ and $1$. The present numerical experiments use the ideal gas equation of state, and the initial temperature is set to maintain the flow $\mathrm{Ma}<0.10$. The initial thermodynamic and flow properties are defined as follows: $\mu_l=0.002$g/(cm.s), $\mu_h=0.006$g/(cm.s), $\rho_l=1.0$g/cm$^3$, $\rho_h=3.0$g/cm$^3$, $\gamma_l=\gamma_h=1.40$, $g=-980$cm/s$^2$, $k_o=10^{-6} \mathrm{cm^2/s^2}$, $S_o=10^{-6} \mathrm{cm}$, and Schmidt and Prandtl numbers equal to one.
%
%
%
%
\subsection{Numerical settings}
\label{sec:2.2}

The CC flow is calculated with the incompressible flow solver ReFRESCO \cite{REFRESCO_CODE_2017}. This is a community based open-usage flow solver optimized for maritime applications. ReFRESCO solves multiphase incompressible viscous flows using the filtered continuity and Navier-Stokes equations, complemented with turbulence closures. The equations are discretized using a finite-volume approach with cell-centered collocated variables, in strong-conservation form, and a pressure-correction equation based on the SIMPLE algorithm is used to ensure mass conservation. Time integration is performed implicitly with a second-order backward scheme. The spacial discretization schemes are also second-order accurate, and we use a second-order upwind based scheme to discretize the convective terms of the governing equations. Four spatio-temporal grid resolutions are used, ranging from $1.04 \times 10^6$ to $4.55 \times 10^6$ cells and time-steps between $8.53\times 10^{-3}$ to $5.21\times 10^{-3}$ time units. These lead to a refinement ratio $r$ of approximately $1.64$. At each implicit time-step, the non-linear system for velocity and pressure is linearised with Picard's method, and a segregated approach is adopted for the solution of all transport equations. The simulations run on double precision, and the iterative convergence criterion \cite{PEREIRA_ACME_2021} guarantees that the maximum normalized residual of all dependent quantities is always smaller than $10^{-5}$.

 The implementation is face-based, which permits grids with elements consisting of an arbitrary number of faces. ReFRESCO was initially developed as a RANS solver \cite{VAZ_OMAE_2009}, where the Reynolds-stress tensor was modeled through linear turbulent viscosity closures. Pereira \cite{PEREIRA_PHD_2018,PEREIRA_OE2_2019} extended the available mathematical models and turbulence closures to PANS, XLES, DDES, IDDES, RANS-EARSM, and RANS-RSM methods. The code is parallelized using MPI and subdomain decomposition, and runs on Linux workstations and HPC clusters. The code is currently being developed, verified and tested at MARIN (the Netherlands) in collaboration with several universities \cite{REFRESCO_CODE_2017}.

The TGV and RT calculations are conducted with the compressible flow solver xRAGE \cite{GITTINGS_CSD_2008}. This code utilizes a finite volume approach to solve the compressible and multi-material conservation equations for mass, momentum, energy, and species concentration. The resulting system of governing equations is resolved through the Harten-Lax-van Leer-Contact \cite{TORO_SW_1994} Riemann solver using a directionally unsplit strategy, direct remap, parabolic reconstruction \cite{COLLELA_JCP_1987}, and the low Mach number correction proposed by Thornber et al. \cite{THORNBER_JCP_2008}. The equations are discretized with second-order accurate methods: the spatial discretization is based on a Godunov scheme, whereas the temporal discretization relies on an explicit Runge-Kutta scheme known as Heun's method. The time-step, $\Delta t$, is defined by prescribing the maximum instantaneous CFL number,
\begin{equation}
\label{eq:2.2_1}
\Delta t = \frac{ \Delta x. \mathrm{CFL}}{3(|V| + c)} \; ,
\end{equation}
where $c$ is the speed of sound, and $\Delta x$ is the grid size. The CFL number is set equal to $0.45$ for the TGV and $0.50$ for the RT. The code can utilize an Adaptive Mesh Refinement (AMR) algorithm for following waves, especially shock-waves and contact discontinuities, but this option is not used in this work to prevent hanging-nodes \cite{PEREIRA_MASTER_2012} and, as such, the simulations use orthogonal uniform hexahedral grids. These have $128^3$, $256^3$, and $512^3$ (and $1024^3$ for $f_k=0.00$) cells for the TGV, and $64^2\times 192$, $128^2\times 384$, $256^2\times 768$ (and $512^2\times 1536$ with CFL$=0.8$ for $f_k=0.00$) cells for the RT. 

xRAGE models miscible material interfaces and high convection-driven flows with a van-Leer limiter \cite{LEER_JCP_1997}, without artificial viscosity, and no material interface treatments \cite{GRINSTEIN_PF_2011,HAINES_PRE_2014}. The solver uses the assumption that cells containing more than one material are in pressure and temperature equilibrium as a mixed cell closure. The effective kinematic viscosity in multi-material problems \cite{PEREIRA_PRE_2021} is defined as
\begin{equation}
\label{eq:2.2_2}
\nu=\sum_{n=1}^{n_t} \nu_n f_n \; ,
\end{equation}
where $n$ is the material index, $n_t$ is the number of materials, and $f_n$ is the volume fraction of material $n$. For the RT flow, the diffusivity ${\cal{D}}$ and thermal conductivity $\kappa$ are defined by imposing the Schmidt ($\mathrm{Sc}\equiv \nu/\cal{D}$) and Prandtl ($\mathrm{Pr}\equiv c_p\mu/\kappa$) numbers equal to one.
%
%
%
\section{Results}
\label{sec:4}
The importance of V\&V to SRS and the advantages of the proposed strategy are illustrated in this section. This is accomplished through its application to the results of the circular cylinder \cite{PEREIRA_IJHFF_2018,PEREIRA_JCP_2018,PEREIRA_OE_2019}, Taylor-Green vortex \cite{PEREIRA_PRF_2021}, and Rayleigh-Taylor \cite{PEREIRA_PRF2_2021,PEREIRA_POF_2021}.
%
%
\subsection{Circular Cylinder}
\label{sec:4.1}

\begin{figure}[t!]
\centering
\subfloat[$\overline{C}_D$.]{\label{fig:4.1_1a}
\includegraphics[scale=0.11,trim=0 0 0 0,clip]{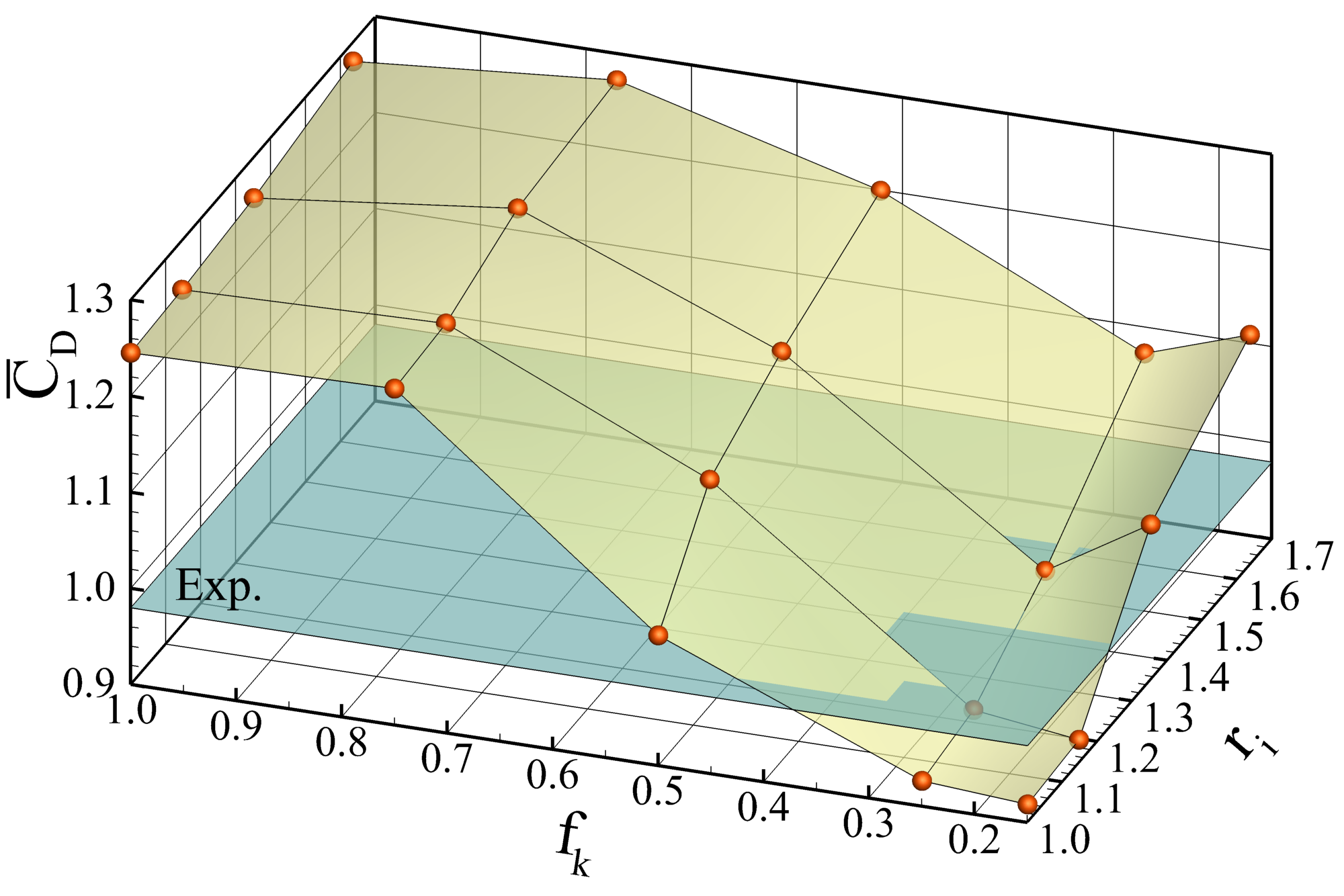}}
~
\subfloat[$C_L'$.]{\label{fig:4.1_1b}
\includegraphics[scale=0.11,trim=0 0 0 0,clip]{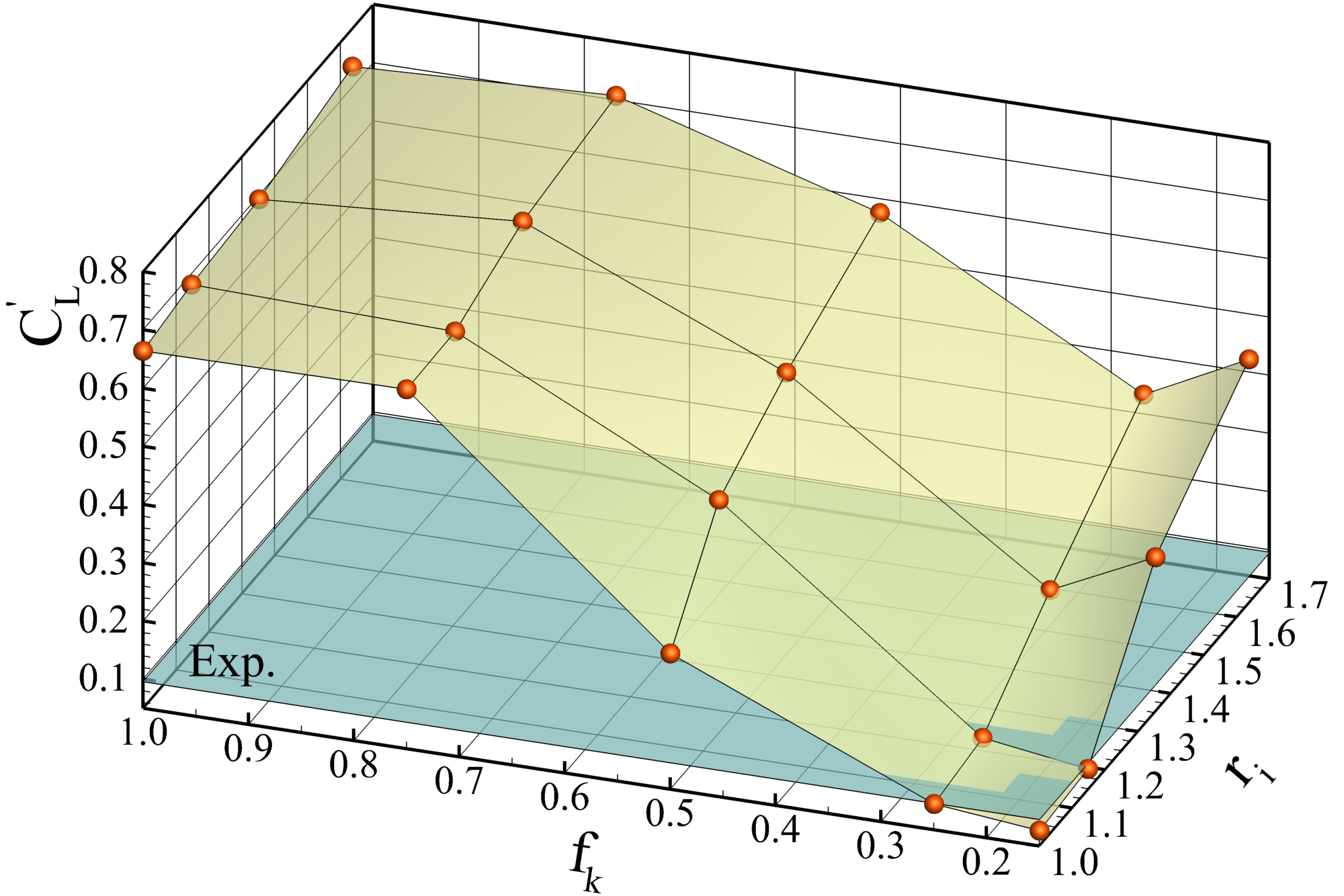}}
\\
\subfloat[$\overline{C}_{pb}$.]{\label{fig:4.1_1c}
\includegraphics[scale=0.11,trim=0 0 0 0,clip]{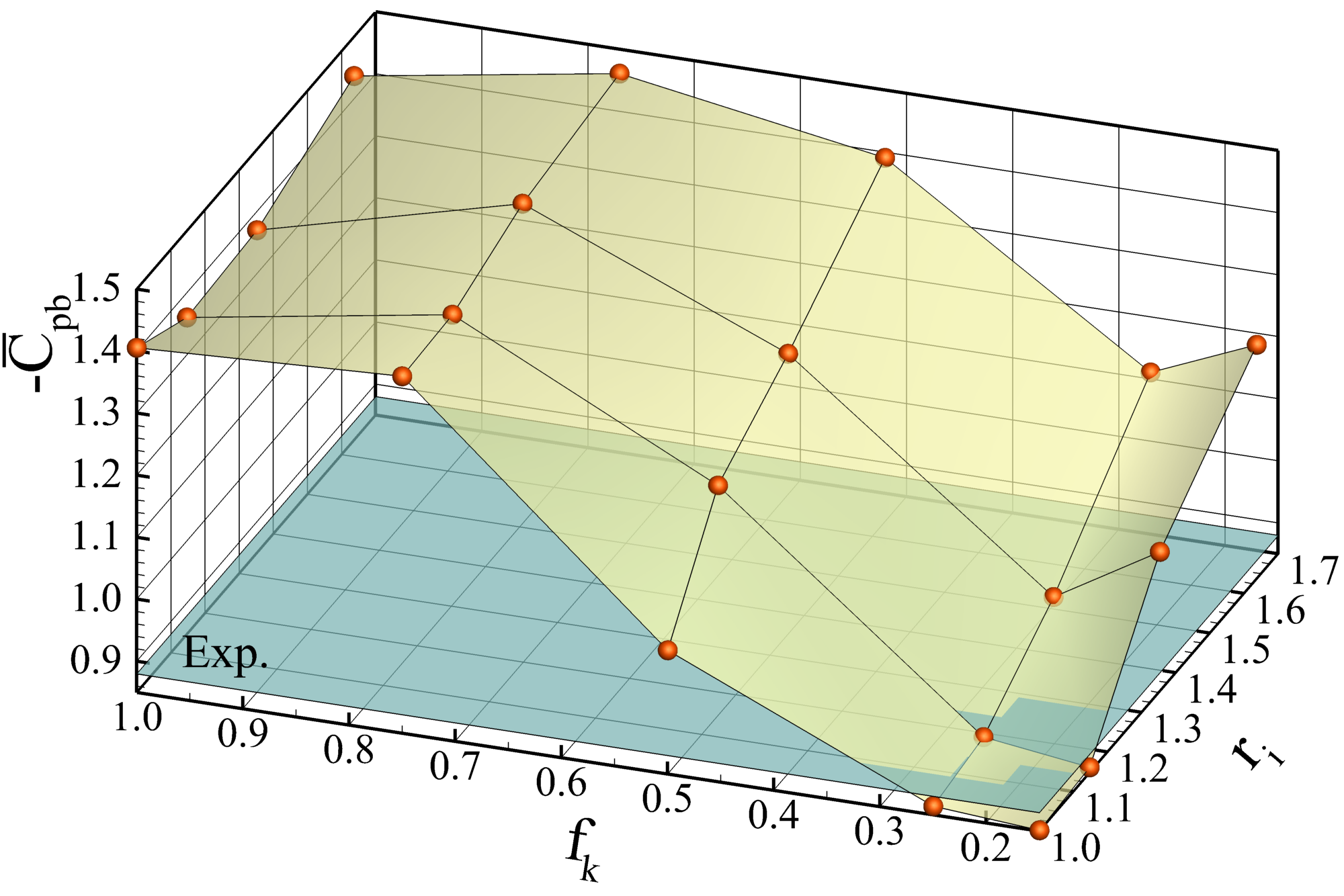}}
~
\subfloat[$\overline{L}_r$.]{\label{fig:4.1_1d}
\includegraphics[scale=0.11,trim=0 0 0 0,clip]{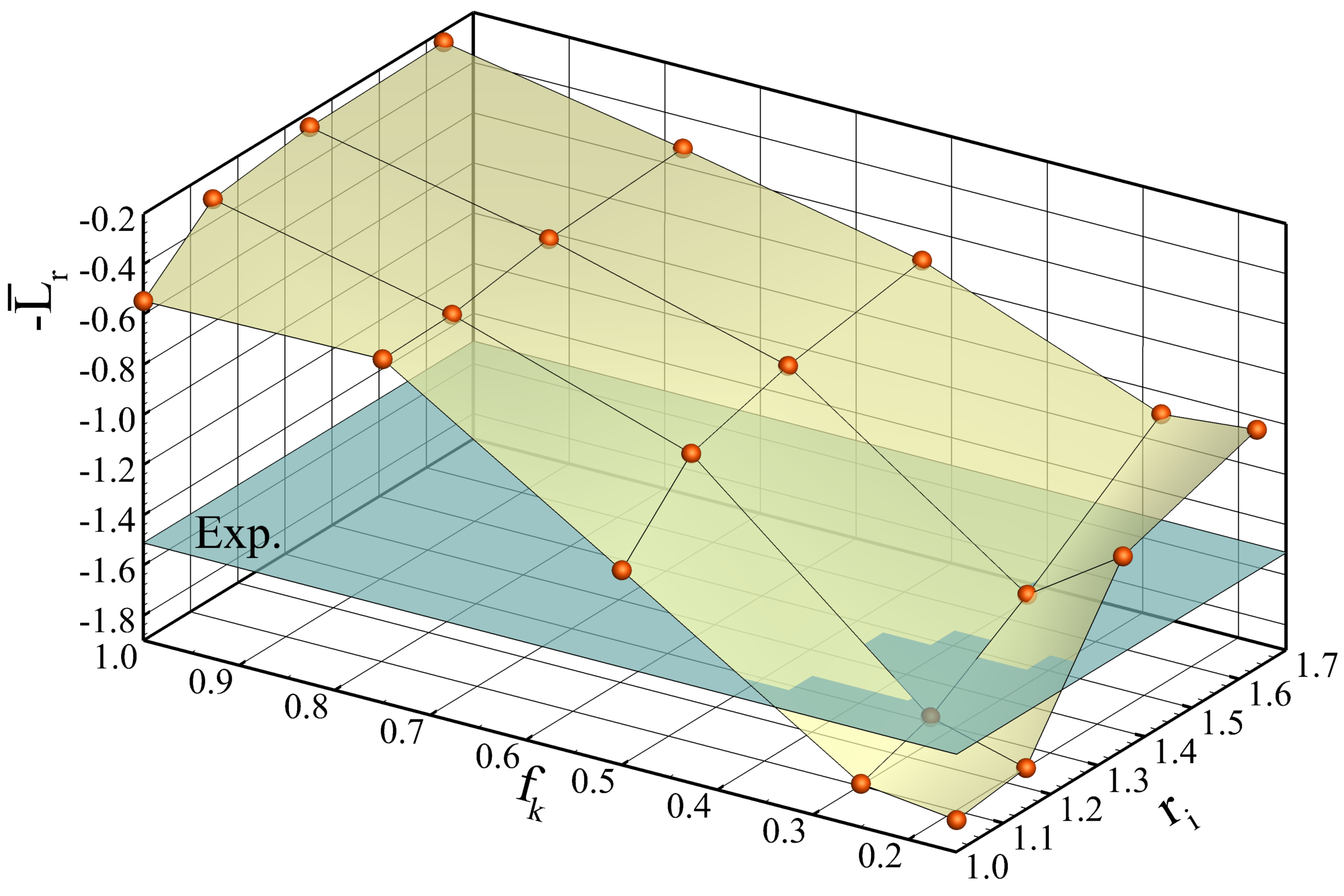}}
\caption{Evolution of $\overline{C}_D$, $C_L'$, $\overline{C}_{pb}$, and $\overline{L}_r$ upon grid, $r_i$, and physical, $f_k$, resolution refinement.}
\label{fig:4.1_1} 
\end{figure}

We initiate the discussion of the results with the flow past a circular cylinder in the sub-critical regime. Figure \ref{fig:4.1_1} presents change with grid, $r_i$, and physical resolution, $f_k$, refinement in three time-averaged quantities: the drag coefficient, $\overline{C}_D$, base pressure  coefficient, $\overline{C}_{pb}$, and recirculation length, $\overline{L}_r$, and one root-mean-square quantity, RMS lift coefficient, $C_L'$. As expected, the simulations converge both with numerical resolution ($r_i\rightarrow 0$) and physical resolution ($f_k\rightarrow 0$).  The results approach the experimental values with varying fidelity depending on the quantity. Yet, it is still possible to observe small discrepancies between the converged numerical results and the experiments. These discrepancies are caused by differences between the numerical and experimental apparatus, i.e., input errors. The experimental flow at the inlet is difficult to replicate because the spectral properties of the turbulence at the inlet were not experimentally measured (only the averaged quantities are available). The length and times scales of disturbances are crucial for transitional problems, and particularly important for numerical predictions of the recirculation region length \cite{FAGE_ARC_1929,SADEH_NASA_1982,NORBERG_JFS_1987,BREUER_FTC_2018}. The differences in aspect ratio (length) of the cylinder \cite{WEST_JFM_1982,SZEPESSY_JFM_1992,NORBERG_JFM_1994}, surface roughness \cite{FAGE_ARC_1929,ACHENBACH_JFM_1971,GUVEN_JFM_1980}, or Re \cite{ESDU_1986} can also have a minor impact to the differences between numerical and experimental measurements. These results reaffirm the importance of the input error, and also show the potential of $r_i$ and $f_k$ refinement studies to create reference solutions that can be used to analyze complex flows. Without all the information in figure \ref{fig:4.1_1}, one could argue that an SRS model at $f_k=0.40$ is ``better'' calculating $\overline{L}_r$ than formulations at finer physical resolutions and draw modeling conclusions based on such idea. However, such a conclusion would be the result of insufficient data, and not properly identifying error cancellation between input and modeling/numerical errors. In this case, the input error decreases $\overline{L}_r$ while the modeling/numerical error increases it. 

Figure \ref{fig:4.1_1} also shows that the selected mathematical model can accurately predict the present flow problem provided an adequate value of $f_k$ and sufficient grid resolution. The results show only small discrepancies between experiments and simulations at $f_k < 0.50$.  This is also seen in the values of $E_c(\phi)$ and $U_v(\phi)$ presented in table \ref{tab:4.1_1} for three $f_k$ (see \cite{PEREIRA_IJHFF_2018,PEREIRA_OE_2019} for the comprehensive analysis). These quantities are obtained using the V\&V20 metric \cite{ASME_BOOK_2009}. For example, the results show that the values of $C_L'$ and $-\overline{C}_{pb}$  decrease from 0.664 and 1.41 at $f_k=1.00$ (RANS) to $0.095$ and 0.86 at $f_k=0.25$, where the experimental values are equal to $0.096$ and $0.86$, respectively. This constitutes a reduction of $E_c(C'_L)$ from $591.8\%$ to $-1.1\%$ and $E_c(\overline{C}_{pb})$ from $59.7\%$ to $-1.8\%$.

We need to emphasize that $C_L'$ is a quantity that does not permit direct comparisons of simulations at different physical resolutions (consistent with the first challenge involved in V\&V of SRS identified in Section \ref{sec:1}). Its magnitude will increase upon $f_k$ refinement, as more of the fluctuations which contribute to $C_L'$ are resolved, and this behavior is independent of the modeling accuracy of the simulation. The present results show that the modeling error of simulations at $f_k=1.00$ is larger than this effect since $C_L'$ grows as $f_k \rightarrow 1.00$.
\begin{table}[t!]
\centering
\caption{Estimated comparison error, $E_c(\phi)$, and validation uncertainty, $U_v(\phi)$, for the time-averaged drag coefficient, $\overline{C}_D$, root-mean-square lift coefficient, $C_L'$, time-averaged base pressure coefficient, $\overline{C}_{pb}$, and recirculation length, $\overline{L}_r$, for different $f_k$. $E_c(\phi)$ and $U_v(\phi)$ are shown as percentage of the experimental value \cite{NORBERG_BBVIV3_2002,NORBERG_JFS_2003,PARNAUDEAU_PF_2008}.}
\label{tab:4.1_1}
\begin{tabular}{C{2.1cm}C{1.2cm}C{1.15cm}C{1.15cm}C{1.15cm}C{1.15cm}C{1.15cm}C{1.15cm}C{1.15cm}}\hline
Model& $\Phi$	   &$\overline{C}_D$	&$C_L'$	&$-\overline{C}_{pb}$	& $\overline{L}_r$ \\\hline
\multirow{3}{*}{\parbox{2cm}{\centering $f_k=1.00$}}  
		&$\phi$		&$1.25$		& $0.664$       	& $1.41$     	&$0.51$     	\\
		&$E_c(\phi)$	&$+27.1$      	&$+591.8$	   	&$+59.7$	 	&$-63.8$      	\\	
		&$U_v(\phi)$	&$\pm1.5$    	&$\pm57.6$ 		&$\pm16.2$  	&$\pm49.7$  	\\\hline
\multirow{3}{*}{\parbox{2cm}{\centering $f_k=0.50$}}
		& $\phi$		& $1.04$		& $0.284$ 		& $1.05$		&$1.12$		\\ 
		&$E_c(\phi)$	&$+5.7$	   	&$+195.7$		&$+19.3$		&$-25.6$		\\
		&$U_v(\phi)$	&$\pm32.2$ 	&$\pm549.7$  	&$\pm61.7$	&$\pm62.0$	\\\hline
\multirow{3}{*}{\parbox{2cm}{\centering $f_k=0.25$}}
           &$\phi$		& $0.93$		& $0.095$		& $0.86$ 		& $1.73$			\\ 
		&$E_c(\phi)$	&$-5.4$		&$-1.1$		&$-1.8$		&$+14.4$			\\
 		&$U_v(\phi)$ 	&$\pm3.9$	&$\pm55.9$	&$\pm6.4$	&$\pm20.2$		\\\hline
\multirow{2}{*}{Exp.}
		&$\phi_e$		&$0.98$		&$0.096$		&$0.88$		&$1.51$			\\
		&$U_e$		&$\pm0.01$		&$\pm0.004$		&$\pm0.01$		&$\pm0.02$			\\\hline
\end{tabular}
\end{table}

A very significant result of figure \ref{fig:4.1_1} from a V\&V perspective is the minima observed on the solutions convergence with $f_k$ on the two coarsest grids ($r_i \ge 1.35$). This behavior occurs at $f_k\le 0.50$ and leads to significantly larger comparison errors at the finest physical resolution ($f_k=0.15$) than at $0.25$. It is caused by insufficient grid resolution to resolve all scales at $f_k=0.15$ and demonstrates that the advantages and potential of SRS models can only be achieved with sufficient numerical resolution. The proposed V\&V strategy identifies this behavior because it clearly distinguishes between the numerical errors and the modeling errors. Importantly, the method does not require reference data to make this distinction.

Another interesting result in figure \ref{fig:4.1_1} is the considerable dependence of solutions near $f_k=0.50$ on the grid resolution (and $f_k$). As observed in table \ref{tab:4.1_1}, the validation uncertainty is highest at $f_k=0.50$. This can be explained in terms of the sensitivity of the flow to the effective Reynolds number, $\mathrm{Re_e}$,
\begin{equation}
\label{eq:4.1_1}
\mathrm{Re_e} \equiv \frac{V_o D}{\nu_e} \; ,
\end{equation}
where $\nu_e$ is an effective viscosity, which can be decomposed as
\begin{equation}
\label{eq:4.1_2}
\nu_e = \nu + \nu_t + \nu_n  \; ,
\end{equation}
with $\nu$, $\nu_t$, $\nu_n$ the molecular, turbulent, and numerical kinematic viscosities, respectively. Note that $\nu_n$ comes from the numerical error. As demonstrated in Pereira et al. \cite{PEREIRA_JCP_2018}, $\mathrm{Re_e}$ can determine the ability of a mathematical model to capture the key instabilities and coherent structures driving this flow. Since $f_k$ dictates $\nu_t$ (and $\nu_n$ indirectly), simulations at values of $f_k$ close to the threshold at which the model begins to resolve the key instabilities and coherent structures of the flow are highly sensitive to $\nu_n$. Alternatively, near a critical physical Reynolds number, the correct physical behavior is recovered only if both the turbulent and numerical viscosities are sufficiently small.  Once again, this can be identified by the proposed V\&V strategy.  These results also represent crucial information for the utilization and development of SRS turbulence models, enabling a better understanding of the predicted flow physics and phenomena which are difficult to model. Finally, it is essential to mention that simulations at $f_k=0.75$ can lead to slightly larger comparison errors than those at $f_k=1.00$. This is a result of the calibration issues discussed in Section \ref{sec:1} that are not overcome by $f_k$ at coarse physical resolutions \cite{PEREIRA_JFE_2019}.
\begin{figure}[t!]
\centering
\subfloat[$\overline{C}_p(\theta)$.]{\label{fig:4.1_2a}
\includegraphics[scale=0.15,trim=0 0 0 0,clip]{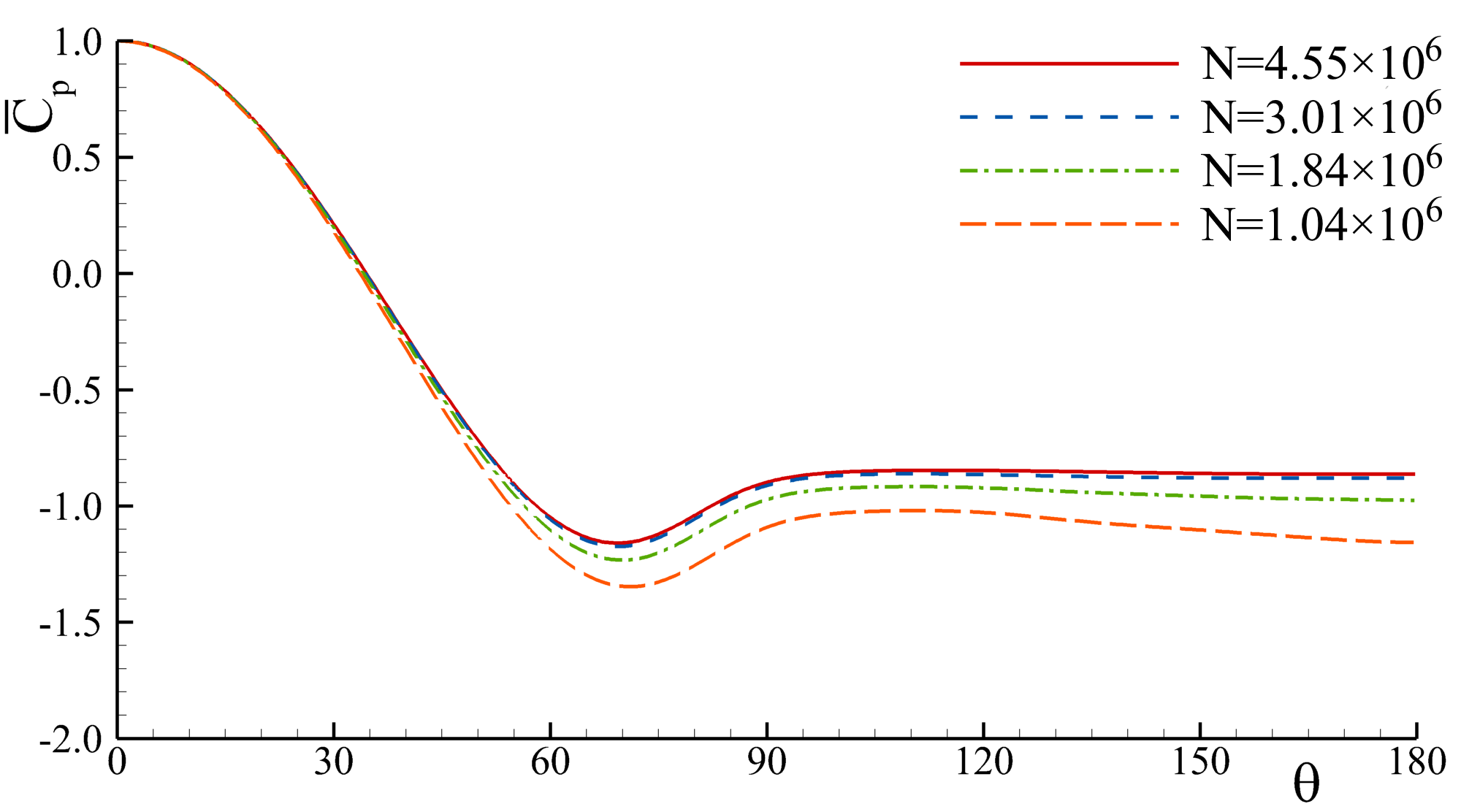}}
~
\subfloat[$\overline{V}_1$.]{\label{fig:4.1_2b}
\includegraphics[scale=0.15,trim=0 0 0 0,clip]{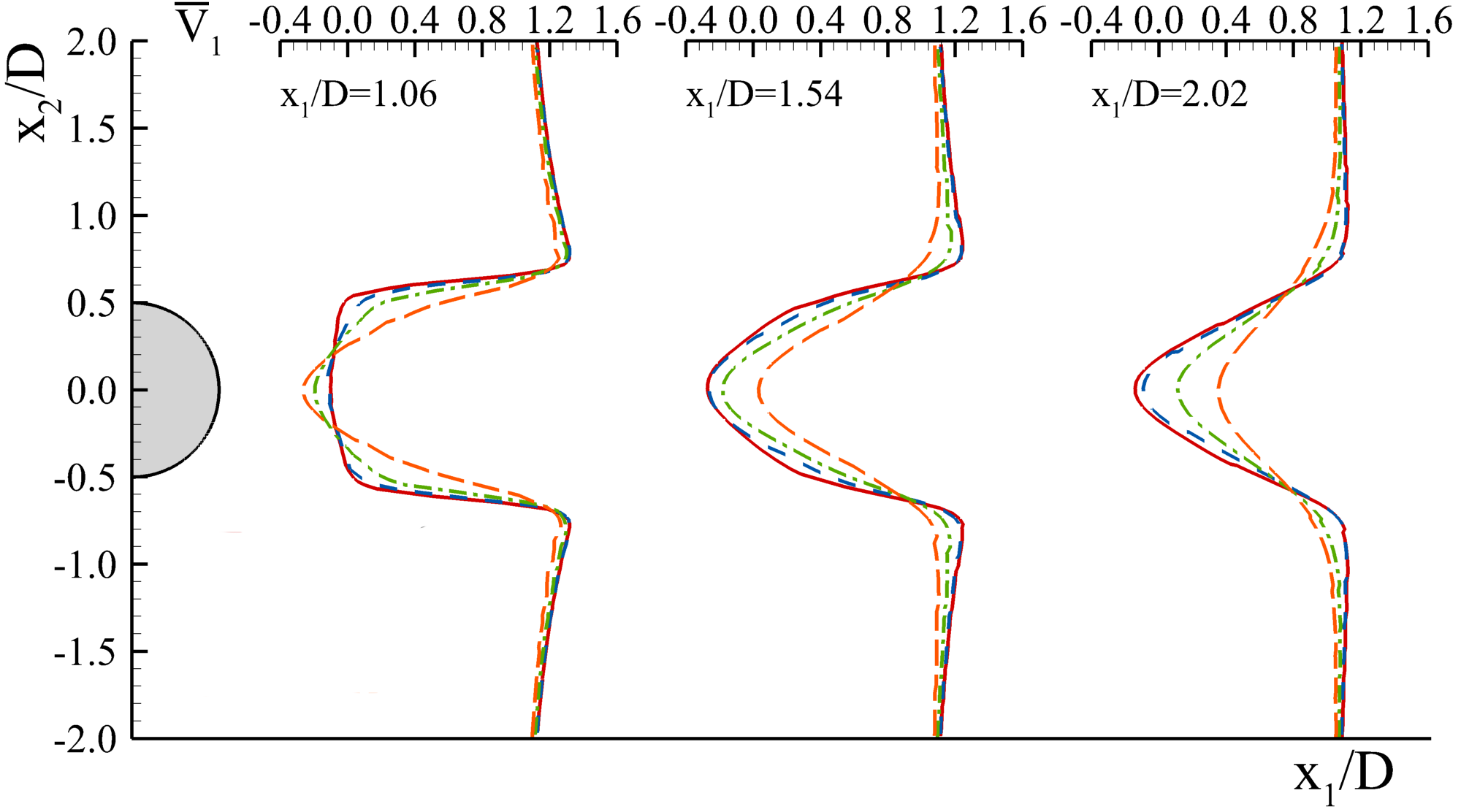}}
\caption{Evolution of $\overline{C}_p(\theta)$ and $\overline{V}_1$ upon grid refinement at $f_k=0.25$.}
\label{fig:4.1_2} 
\end{figure}
\begin{figure}[t!]
\centering
\subfloat[$\overline{C}_p(\theta)$.]{\label{fig:4.1_3a}
\includegraphics[scale=0.15,trim=0 0 0 0,clip]{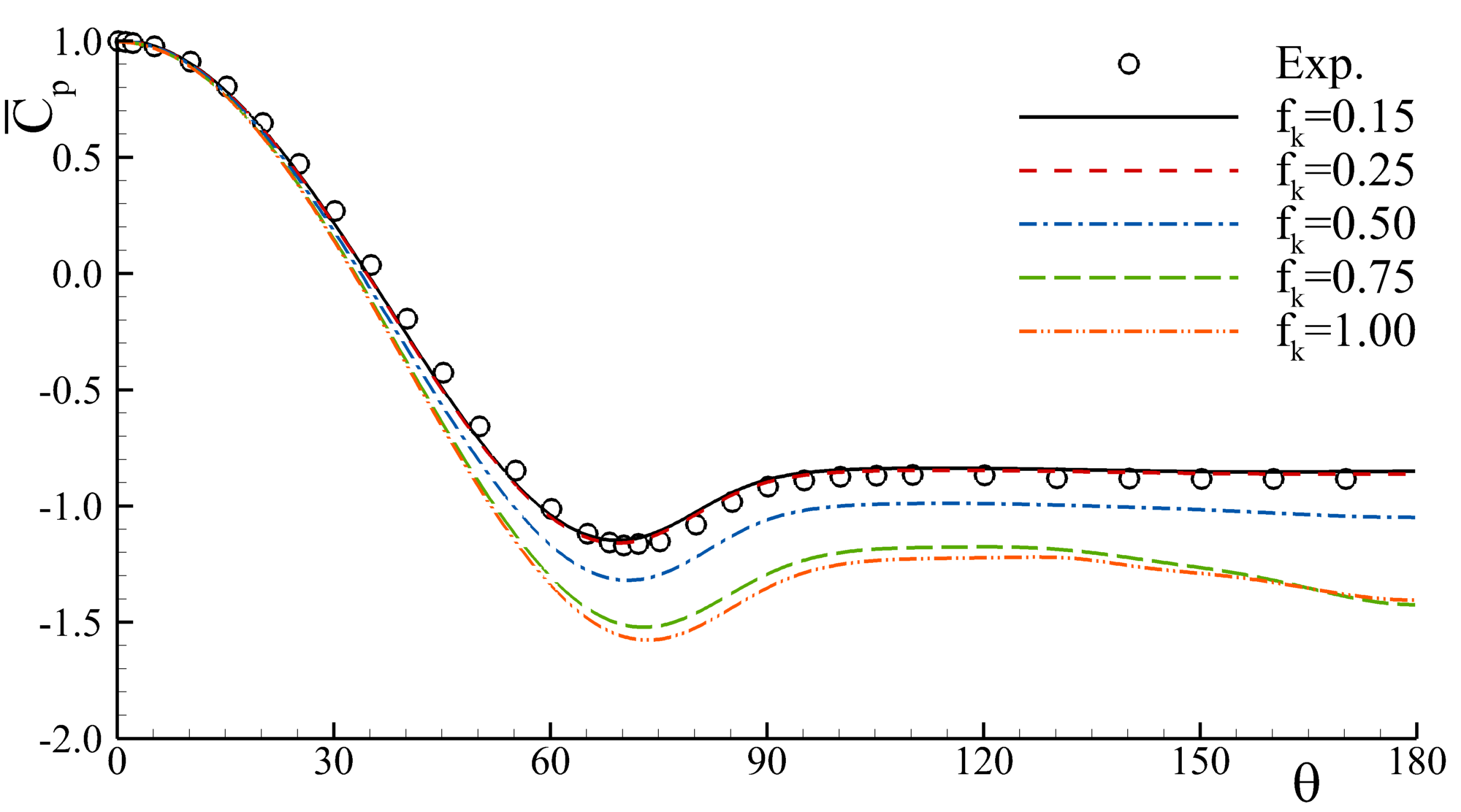}}
~
\subfloat[$\overline{V}_1$.]{\label{fig:4.1_3b}
\includegraphics[scale=0.15,trim=0 0 0 0,clip]{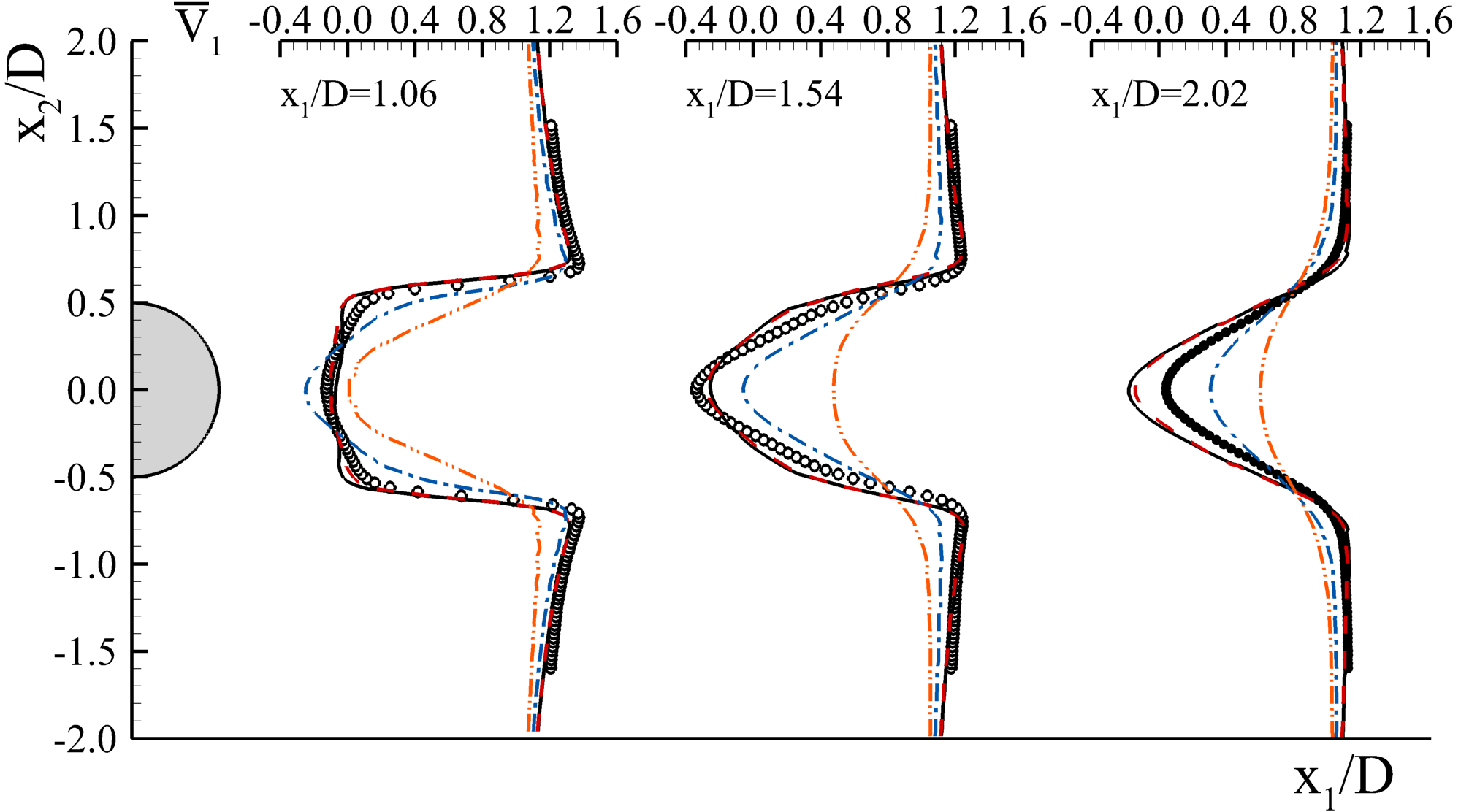}}
\caption{Evolution of $\overline{C}_p(\theta)$ and $\overline{V}_1$ upon physical resolution refinement on grid $g_1$.}
\label{fig:4.1_3} 
\end{figure}

The trends of figure \ref{fig:4.1_1} and table \ref{tab:4.1_1} are also observed in predictions of local flow quantities. This is shown in the following figures, which show the  time-averaged pressure coefficient on the cylinder's surface, $\overline{C}_p(\theta)$, and the streamwise velocity profiles, $\overline{V}_1$, in the near-wake at different grid resolutions for constant $f_k=0.25$ (figure \ref{fig:4.1_2}) and different values of $f_k$ using the finest grid resolution (figure \ref{fig:4.1_3}). All solutions converge upon the grid and physical resolution refinement and tend to the experimental measurement. Once again, the differences between simulations at different physical resolutions on the finest grid depend on the ability of the model to accurately resolve the instabilities driving the flow physics at a given $f_k$. Such behavior is also observed for other SRS formulations \cite{PEREIRA_JFE_2019,PEREIRA_OE_2019}. 
%
%
%
\subsection{Taylor-Green Vortex}
\label{sec:4.2}

\begin{figure}[t!]
\centering
\subfloat[$t=5$.]{\label{fig:4.2_1a}
\includegraphics[scale=0.10,trim=0 0 0 0,clip]{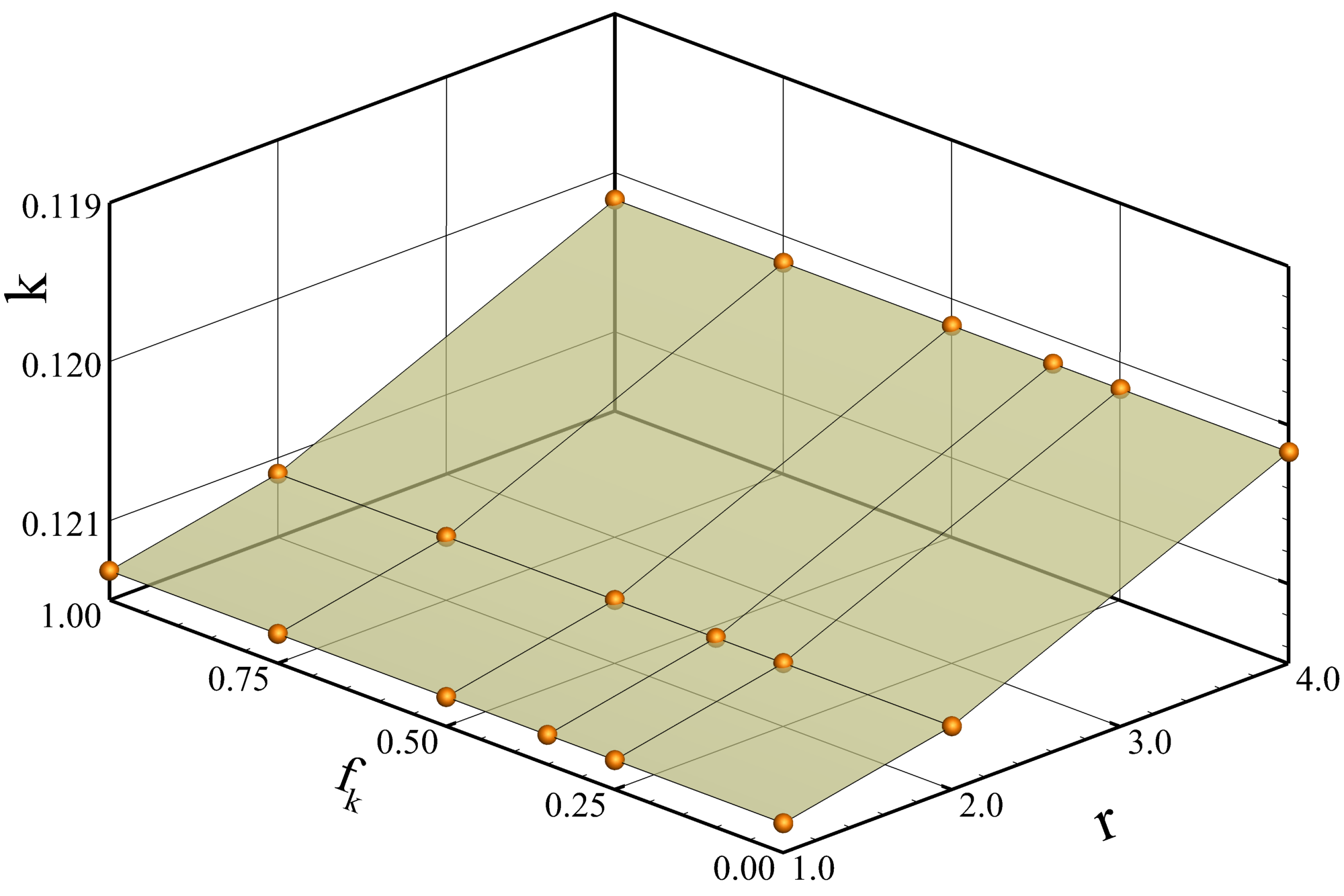}}
~
\subfloat[$t=7$.]{\label{fig:4.2_1b}
\includegraphics[scale=0.10,trim=0 0 0 0,clip]{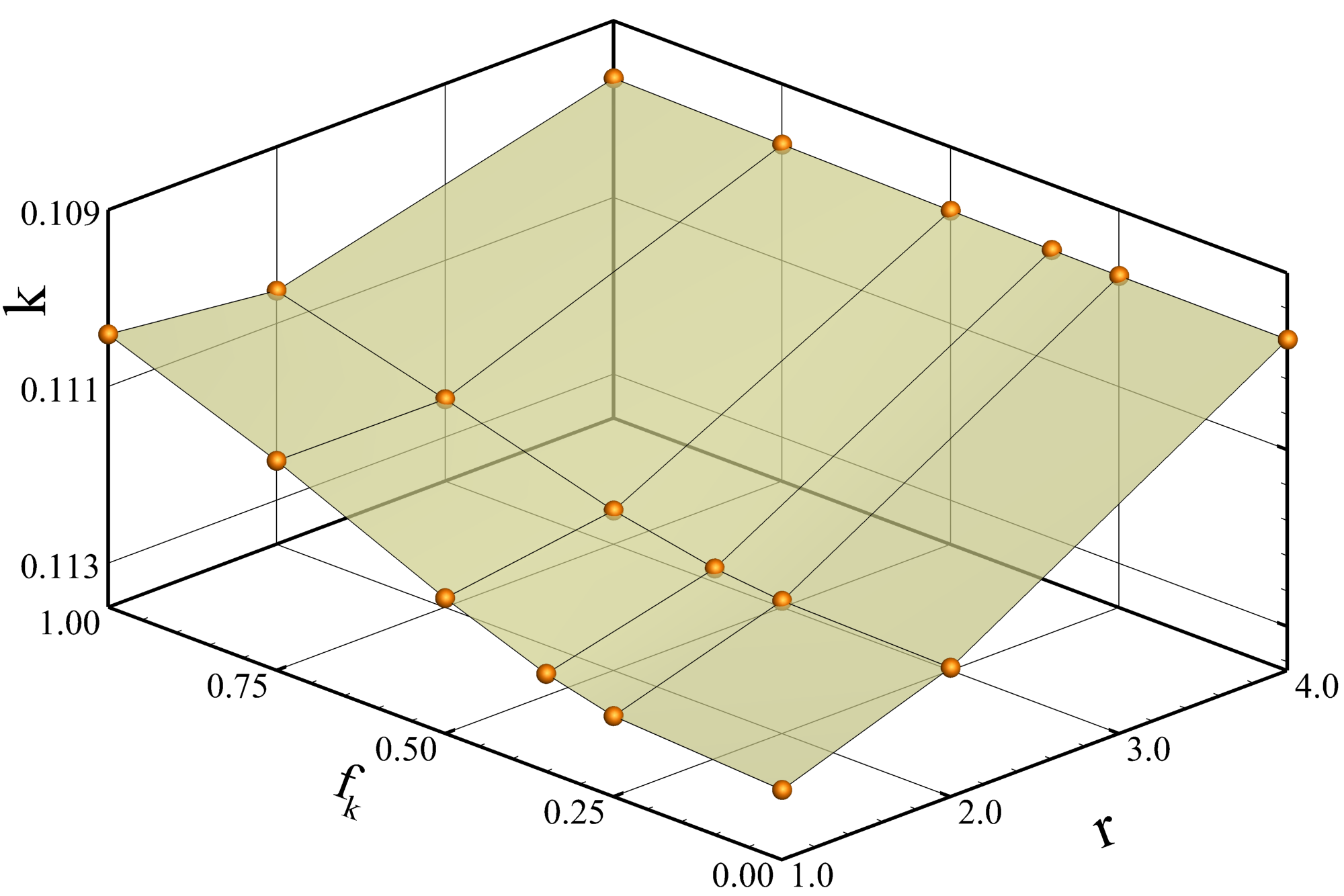}}
\\
\subfloat[$t=9$.]{\label{fig:4.2_1c}
\includegraphics[scale=0.10,trim=0 0 0 0,clip]{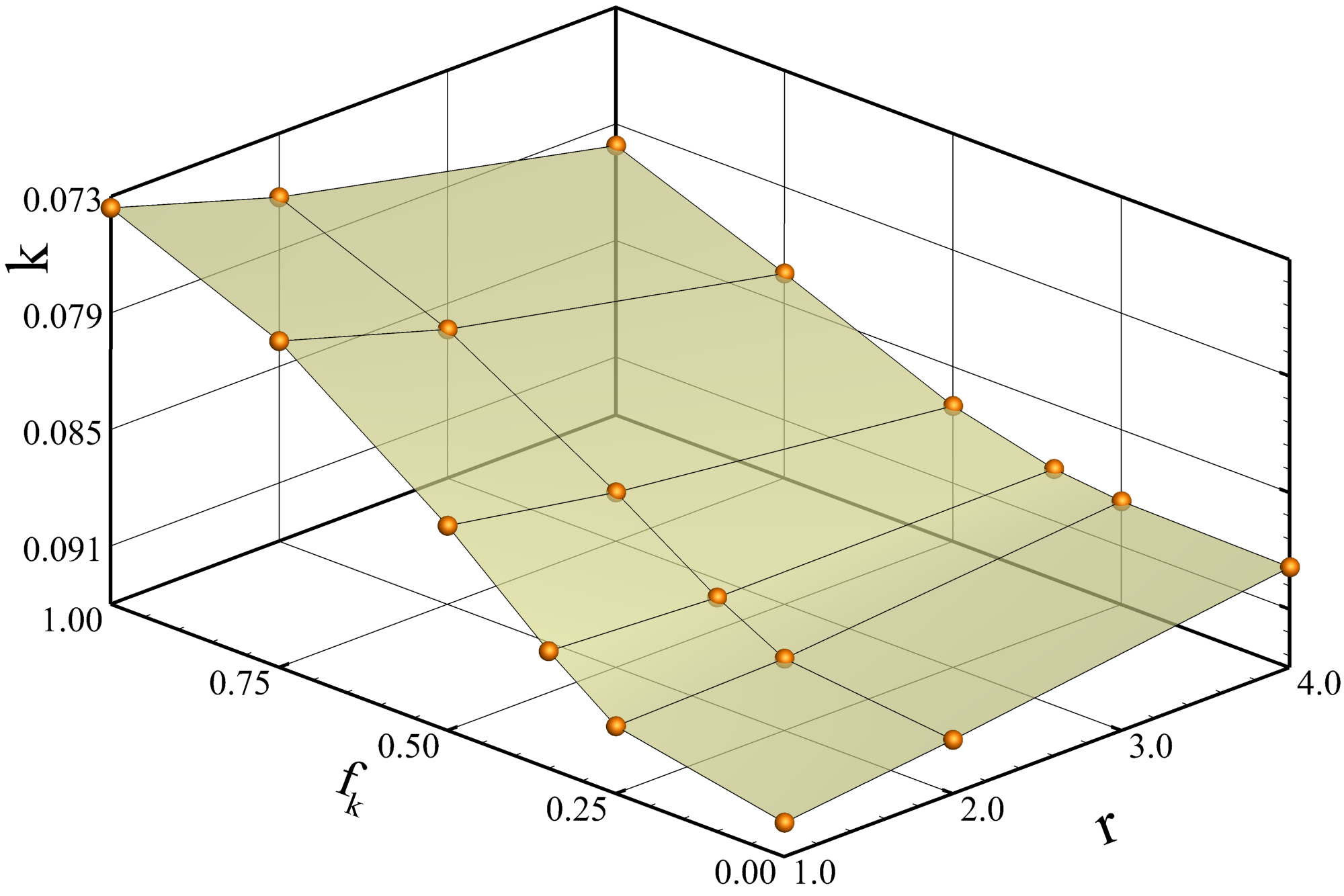}}
~
\subfloat[$t=11$.]{\label{fig:4.2_1d}
\includegraphics[scale=0.10,trim=0 0 0 0,clip]{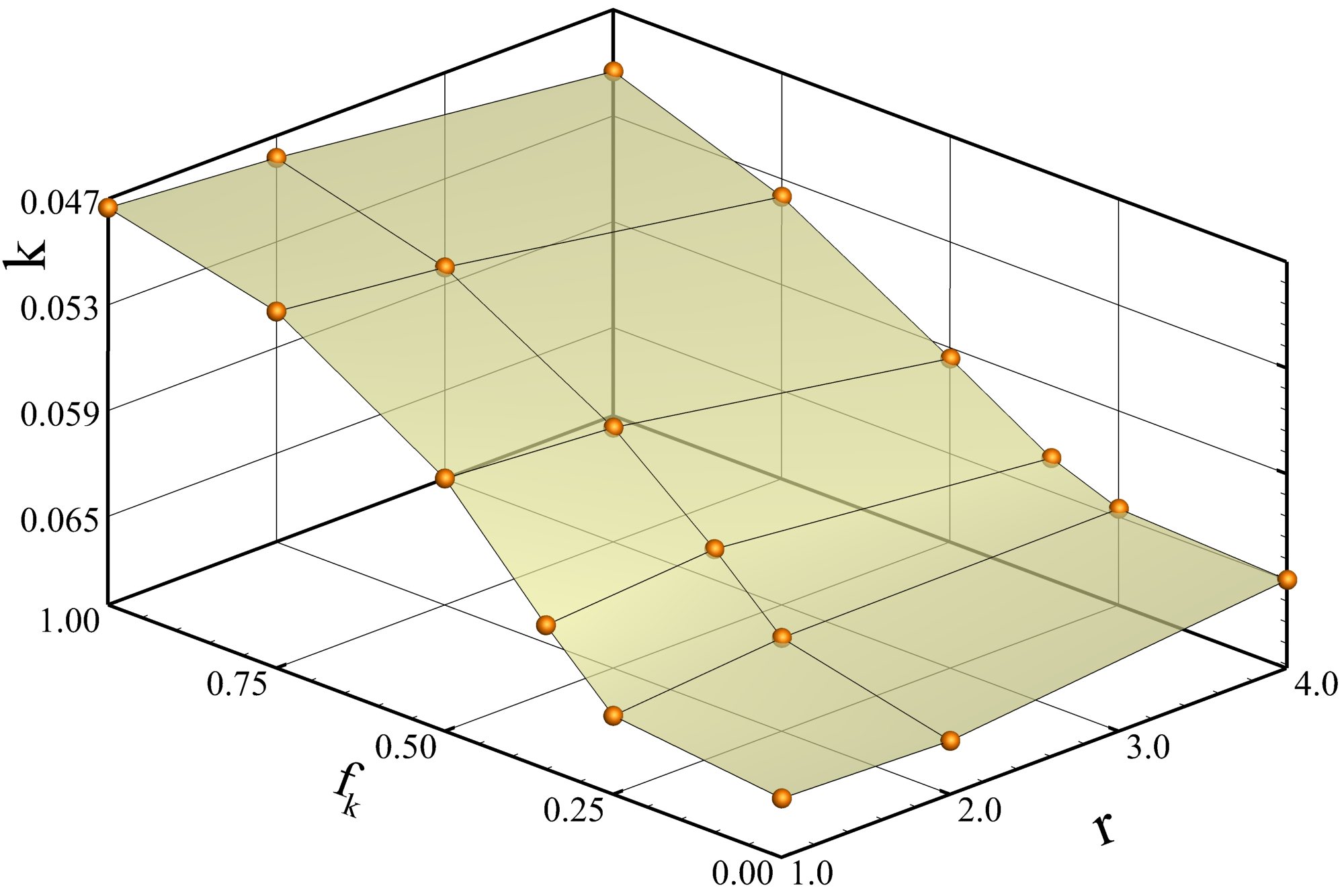}}
\caption{Temporal evolution of the total kinetic energy, $k$, with the grid and $f_k$.}
\label{fig:4.2_1} 
\end{figure}

Next, we analyze the TGV results. This is a transient problem and, as such, V\&V exercises are significantly more challenging and highly dependent on time. A flow misrepresentation at instant $t$ will affect the subsequent instants and can compromise the overall simulation. This renders the accurate simulation and evaluation of the TGV (and RT) flow more complex than for the CC.

Figure \ref{fig:4.2_1} depicts the variation of the total (resolved and modeled) kinetic energy, $k$, with the grid, $r_i$, and physical, $f_k$, resolution at instants before, during, and after the onset of turbulence (the solution at $f_k=0$ on $N=1024^3$ is not included). As discussed in Pereira et al. \cite{PEREIRA_PRF_2021}, the period  $t\leq 12$ is crucial to the flow dynamics and simulation accuracy. At $t=5$, all simulations converge upon grid refinement ($r_i \rightarrow 0$) and are independent of $f_k$. This stems from the fact that the flow is laminar so that all simulations predict negligible magnitudes of turbulent stresses \cite{PEREIRA_PRF_2021}. Later at $t=7$, when the flow is supposed to undergo transition to turbulence, the computations become dependent on $f_k$ and $r_i$. Yet, it is observed that the solutions converge upon the refinement of these parameters, and the discrepancies between simulations at $f_k\leq 0.35$ are small. Also, it is possible to infer that varying physical and numerical resolution cause similar variations on the solutions of $k$. At later times, the impact of $f_k$ relative to $r_i$ grows significantly, but the solutions still converge upon $f_k$ and $r_i$ refinement.

\begin{figure}[t!]
\centering
\subfloat[$f_k=1.00$.]{\label{fig:4.2_2a}
\includegraphics[scale=0.075,trim=0 0 0 0,clip]{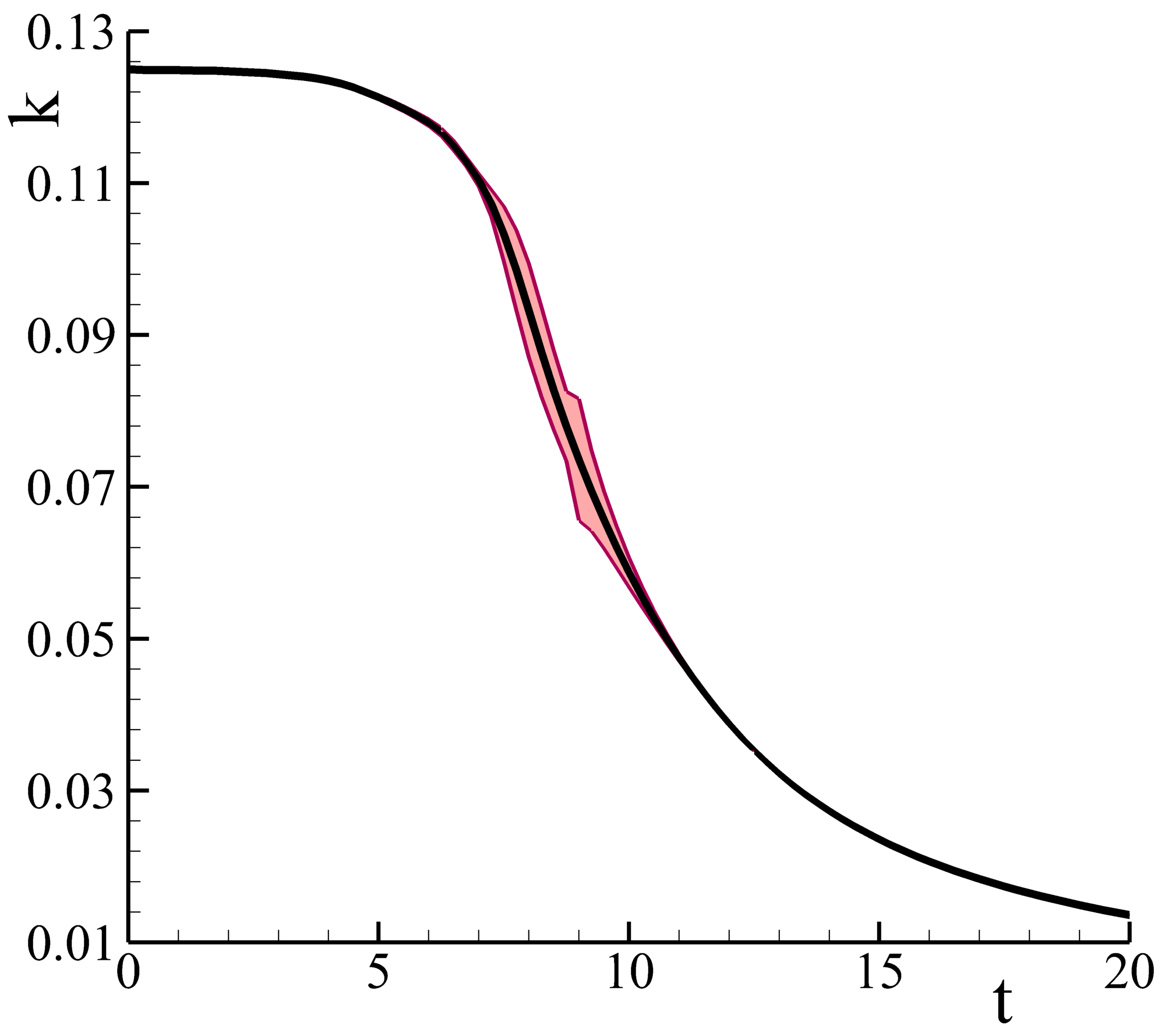}}
~
\subfloat[$f_k=0.50$.]{\label{fig:4.2_2b}
\includegraphics[scale=0.075,trim=0 0 0 0,clip]{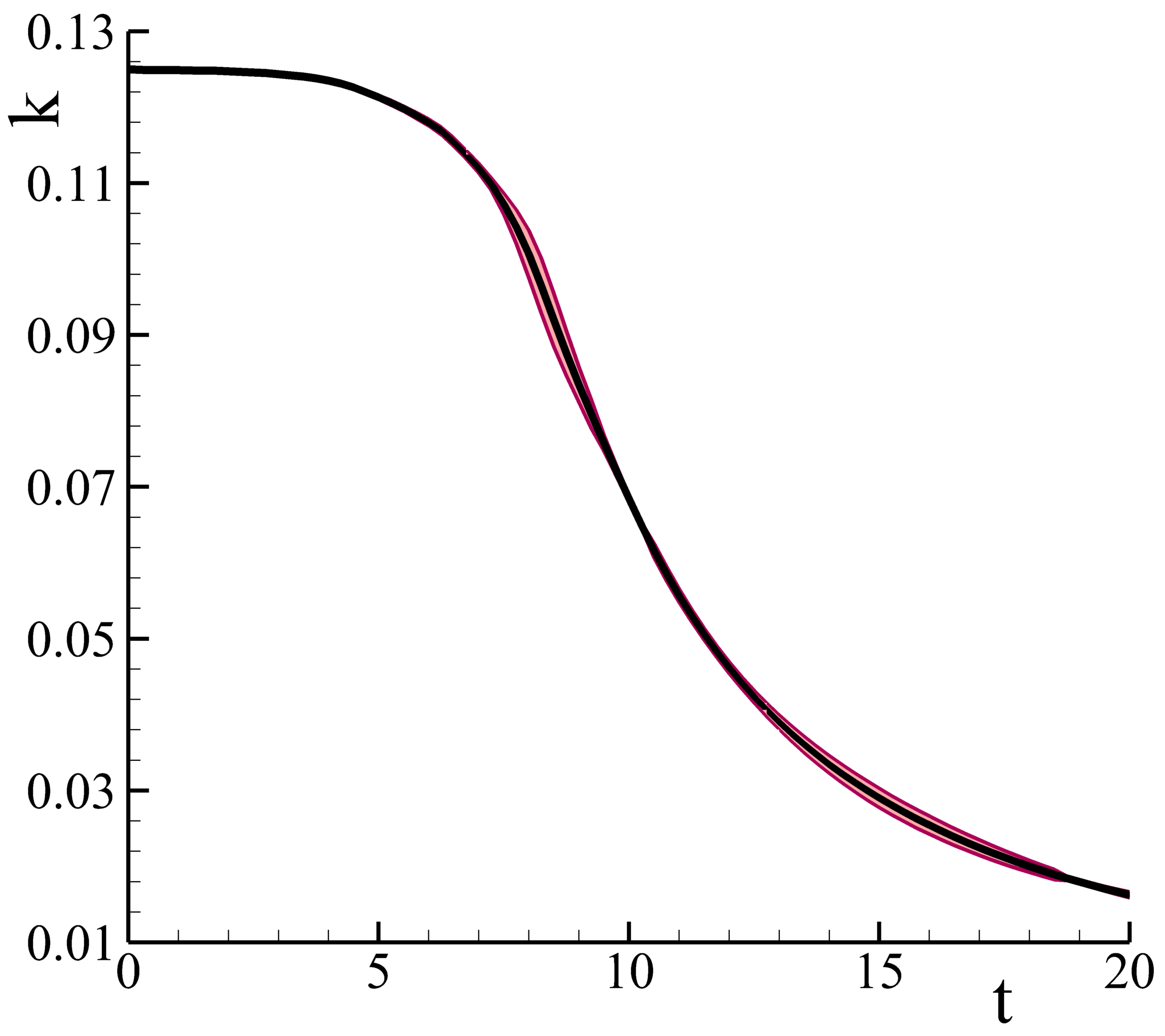}}
~
\subfloat[$f_k=0.35$.]{\label{fig:4.2_2c}
\includegraphics[scale=0.075,trim=0 0 0 0,clip]{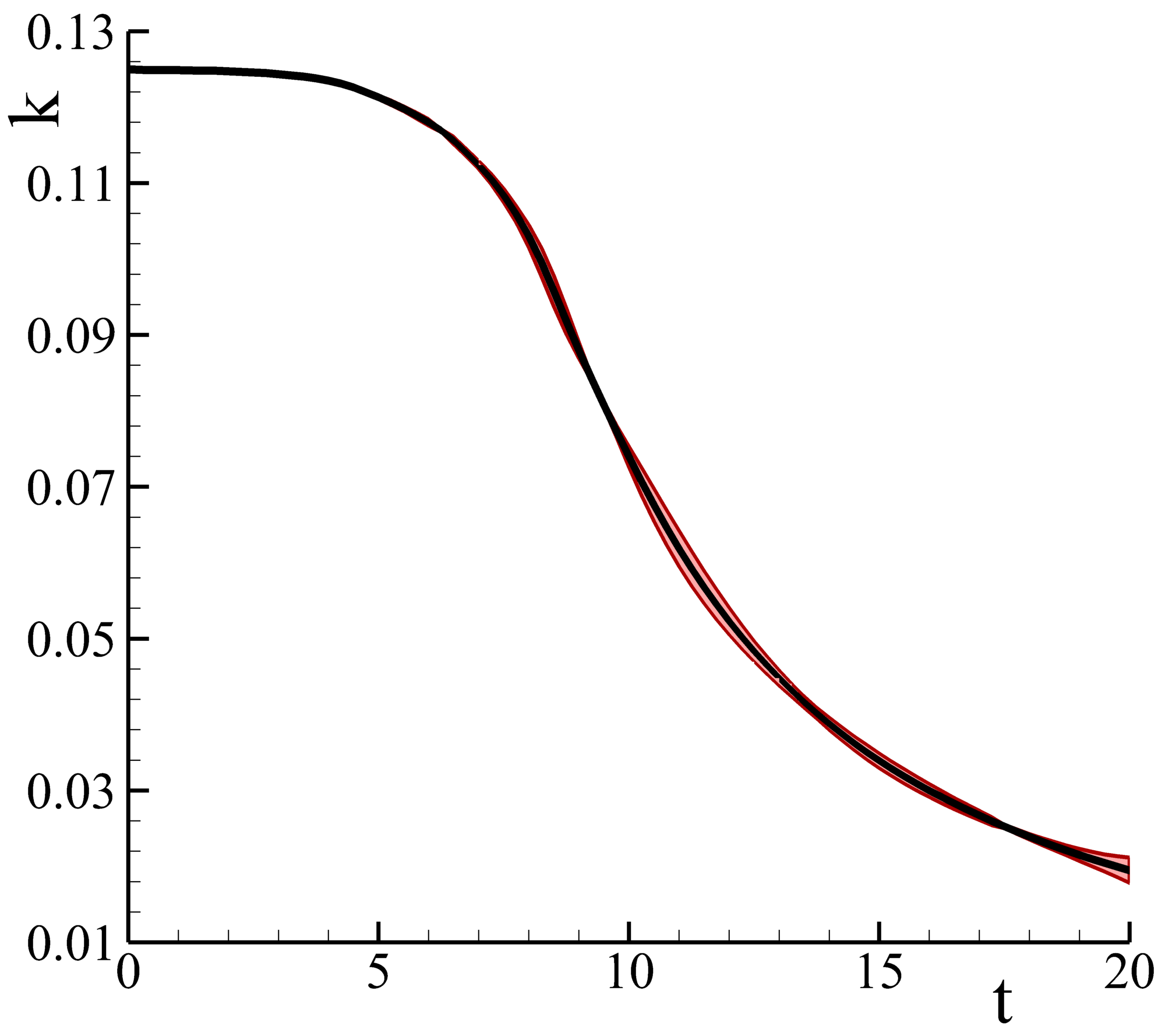}}
\\
\subfloat[$f_k=0.25$.]{\label{fig:4.2_2d}
\includegraphics[scale=0.075,trim=0 0 0 0,clip]{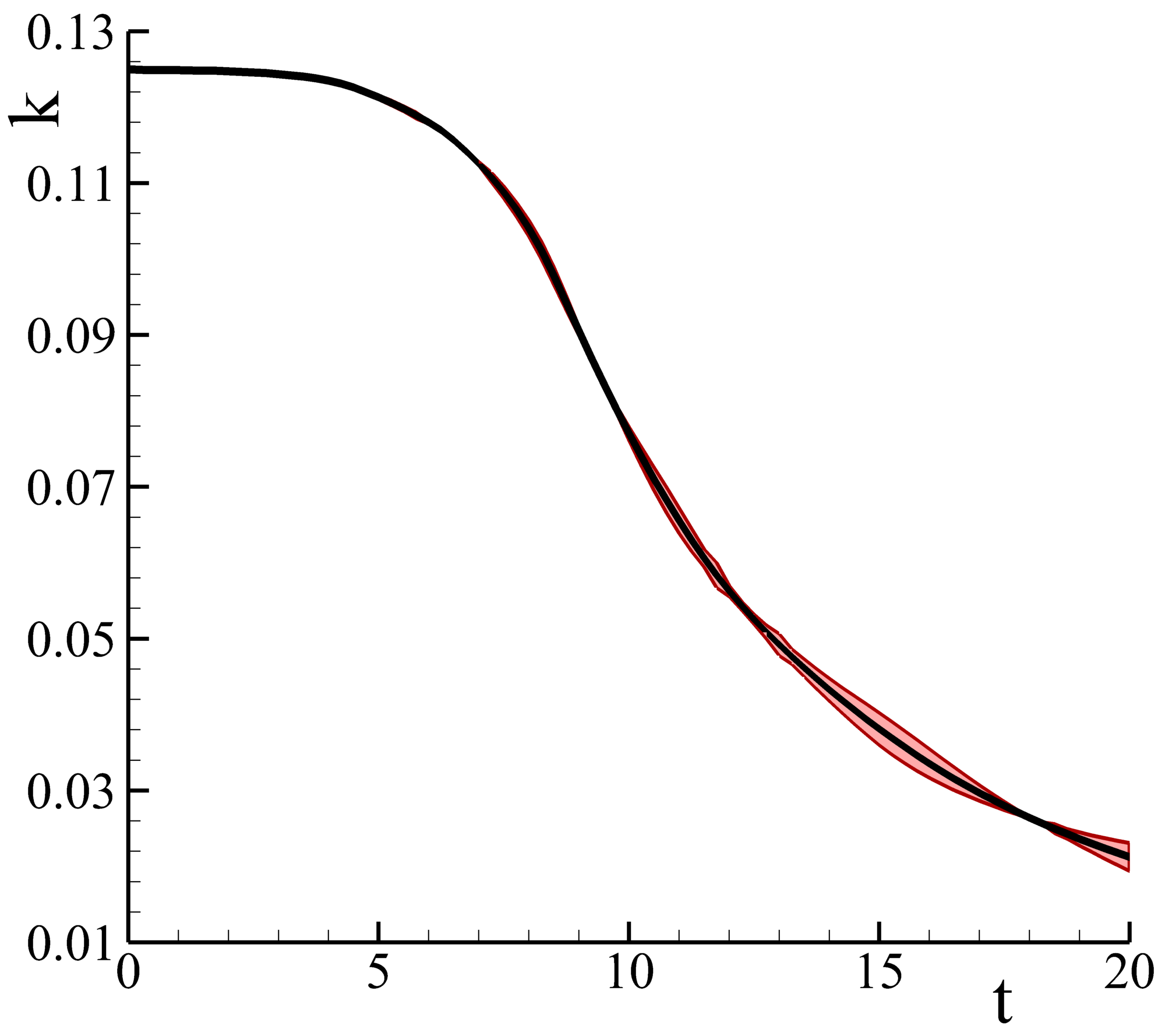}}
~
\subfloat[$f_k=0.00$.]{\label{fig:4.2_2e}
\includegraphics[scale=0.075,trim=0 0 0 0,clip]{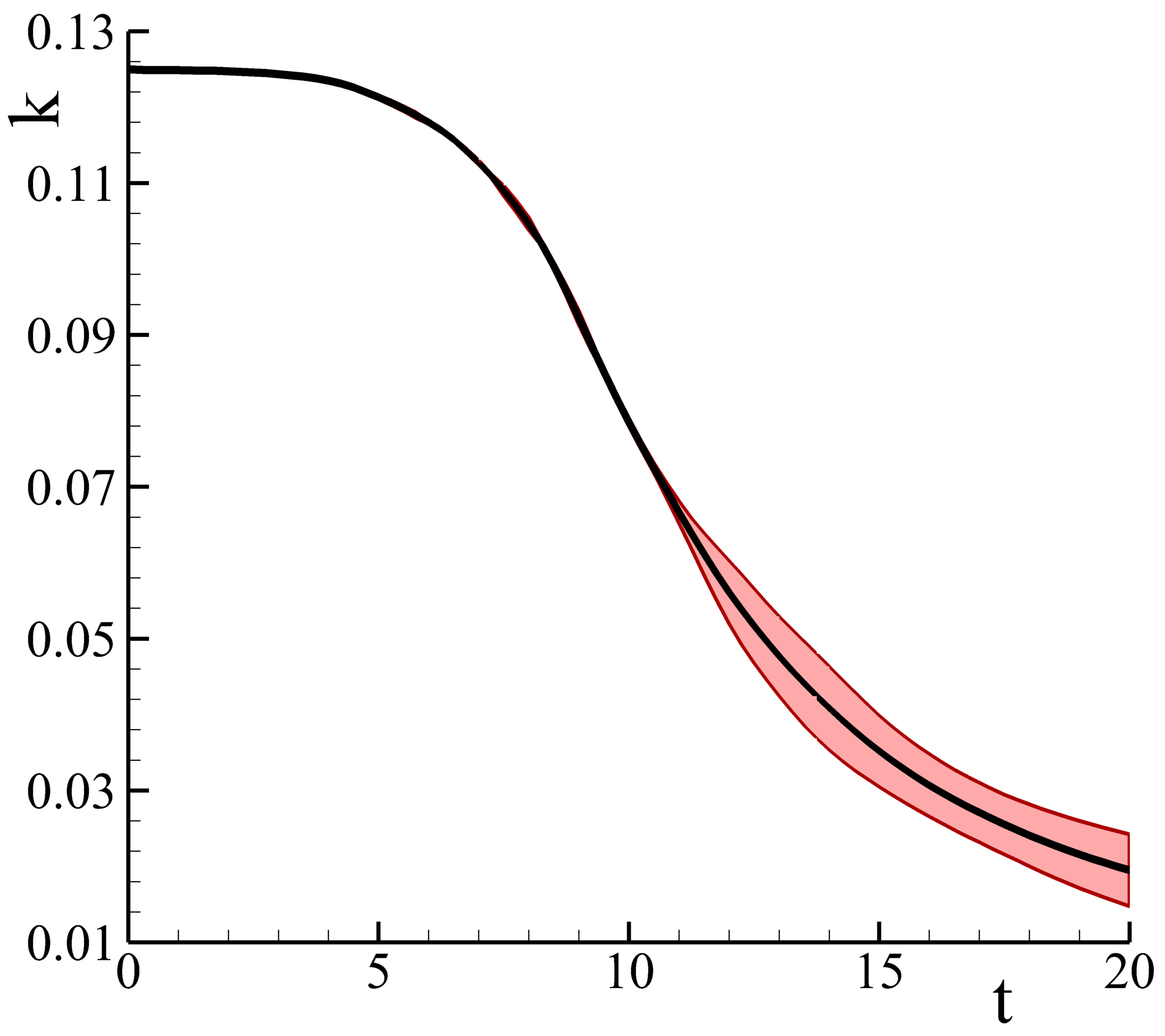}}
~
\subfloat[$f_k=0.00$ on $N=1024^3$.]{\label{fig:4.2_2f}
\includegraphics[scale=0.075,trim=0 0 0 0,clip]{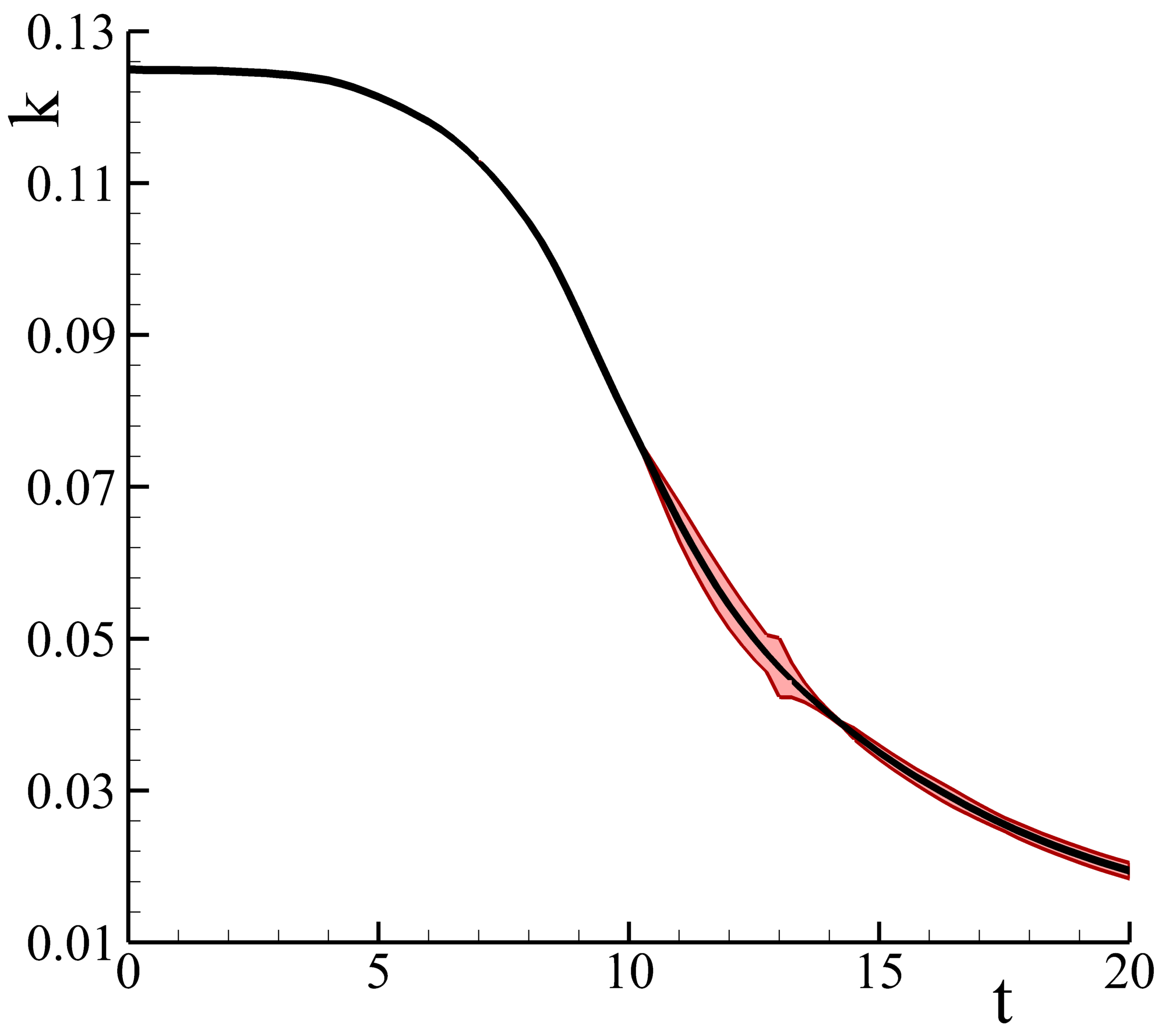}}
\caption{Temporal evolution of the total kinetic energy numerical uncertainty, $U_n(k)$, for different $f_k$.}
\label{fig:4.2_2} 
\end{figure}

Figure \ref{fig:4.2_2} presents the temporal evolution of $k$ and respective numerical uncertainty, $U_n(k)$, at different physical resolutions, $f_k$. $U_n(k)$ is estimated with the method proposed by E\c{c}a and Hoekstra \cite{ECA_JCP_2014} using the three finest grids available: $N=128^3$, $256^3$, and $512^3$ for all $f_k$, and $N=256^3$, $512^3$, and $1024^3$ for $f_k=0.00$. In figure \ref{fig:4.2_2}, the black line indicates $k(t)$, and the height of the red area the value of $U_n(k)$ at a given instant. We assume the discretization error as the main contributor to the numerical error. It is important to reiterate that $U_n(k)$ is the uncertainty associated with the numerical solution procedure and so the red envelope in figure \ref{fig:4.2_2} is the range in which we expect the exact solution to the physical model to fall. The comparison error, $E_c(k)$, which is the difference between the numerical solution and the reference solution, can be large or small, independent of the numerical uncertainty. For example, for $f_k=1$, the error is large (figure \ref{fig:4.2_1}), because the model performs poorly, but the numerical uncertainty at late time can be small, if the numerical solution is well converged.  This means the numerical solution accurately represents the mathematical model, but that the model does not accurately capture the physics. Conversely, for $f_k=0$, at late time, the modeling error is small, but the numerical uncertainty is expected to be larger (for the same grid) because of the high requirements for both numerical and statistical convergence of a fully resolved turbulent flow. For a better physical and numerical interpretation of the results, recall that the onset of turbulence is expected to occur at $t\approx 7$.

Referring to the simulations at $f_k\ge 0.50$, the results exhibit reduced $U_n(k)$ for most of the simulation time. Taking the case at $f_k=1.00$ (RANS), the estimates of $U_n(k)$ do not exceed $3.3\%$ of the predicted $k$. The exception lies in the period between $t=6$ and $10$, in which $U_n(k)$ can reach $10.8\%$. Such values of $U_n(k)$ are mostly caused by the difficulties of  predicting the onset of turbulence with a RANS closure \cite{PEREIRA_PRF_2021}. RANS models overpredict turbulence during transition, leading to a more rapid dissipation of $k$, which increases the flow gradients steepness and, consequently, the simulations' local numerical requisites. This illustrates the importance of a careful numerical and modeling interpretation of V\&V exercises. 

On the other hand, the refinement of $f_k$ to $0.35$ and $0.25$ reduces the predictions' numerical uncertainty significantly. $U_n(k)$ is negligible at early times and does not exceed $8.7\%$ at late times when the flow features high-intensity turbulence and $k$ is significantly smaller than at $t=0$. However, the further refinement of physical resolution ($f_k=0.0$) increases $U_n(k)$ at $t>11.0$ considerably.  This is the period when the flow is characterized by high-intensity turbulence and the largest numerical requirements since $f_k=0.00$ aims to resolve all flow scales. For this reason, for simulations at $f_k=0.00$, even the grid with $N=1024^3$ cells still has higher numerical uncertainties than those at $f_k=0.25$ on $N=512^2$. The simulation at $f_k=0.00$ on the finest grid leads to $(U_n(k))_{\max}=8.4\%$ at $t=13$, whereas $U_n(k)$ does not exceed $2.9\%$ at this instant for $f_k=0.25$.

\begin{figure}[t!]
\centering
\subfloat[$k$.]{\label{fig:4.2_3a}
\includegraphics[scale=0.11,trim=0 0 0 0,clip]{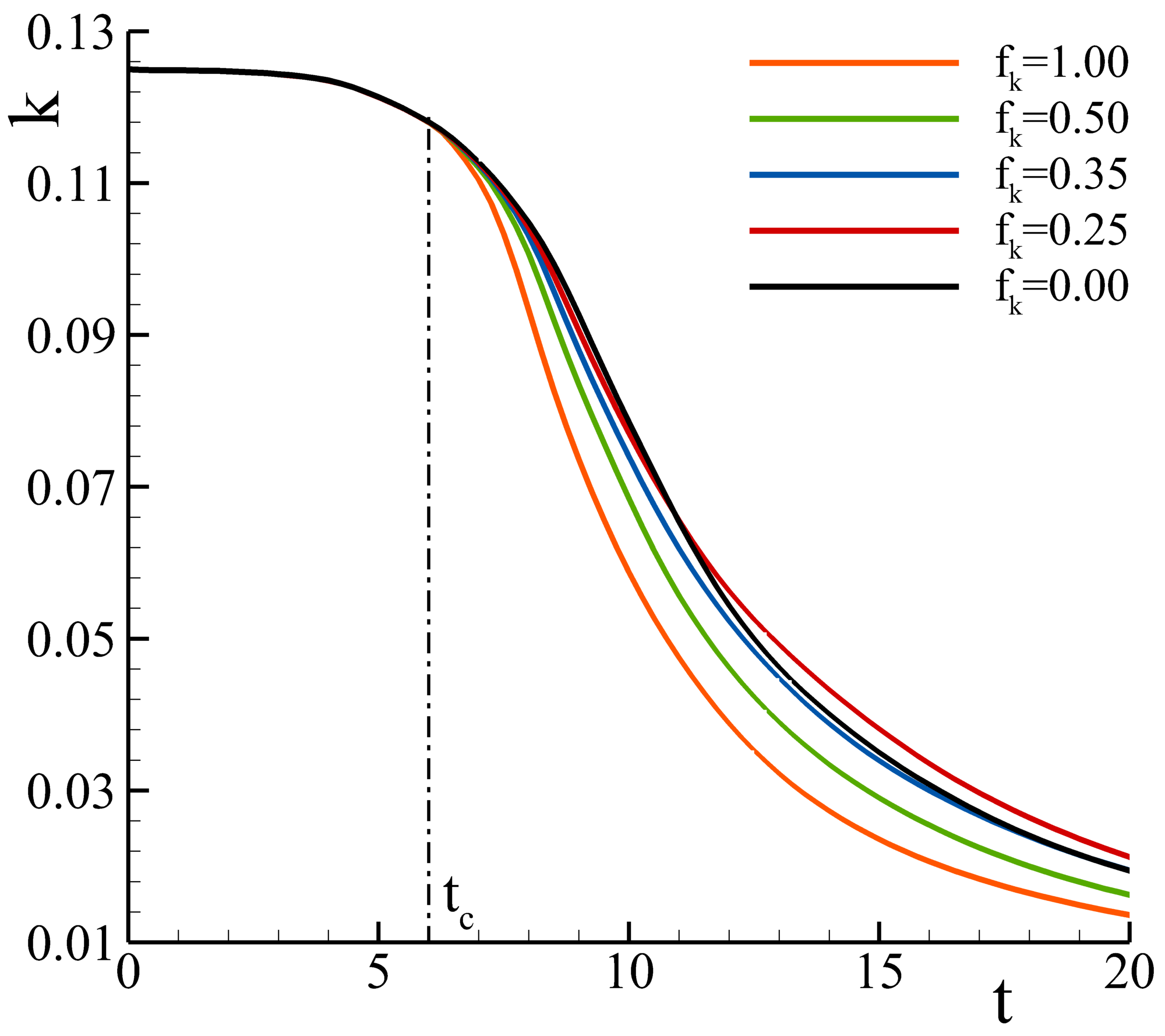}}
~
\subfloat[$\varepsilon$.]{\label{fig:4.2_3b}
\includegraphics[scale=0.11,trim=0 0 0 0,clip]{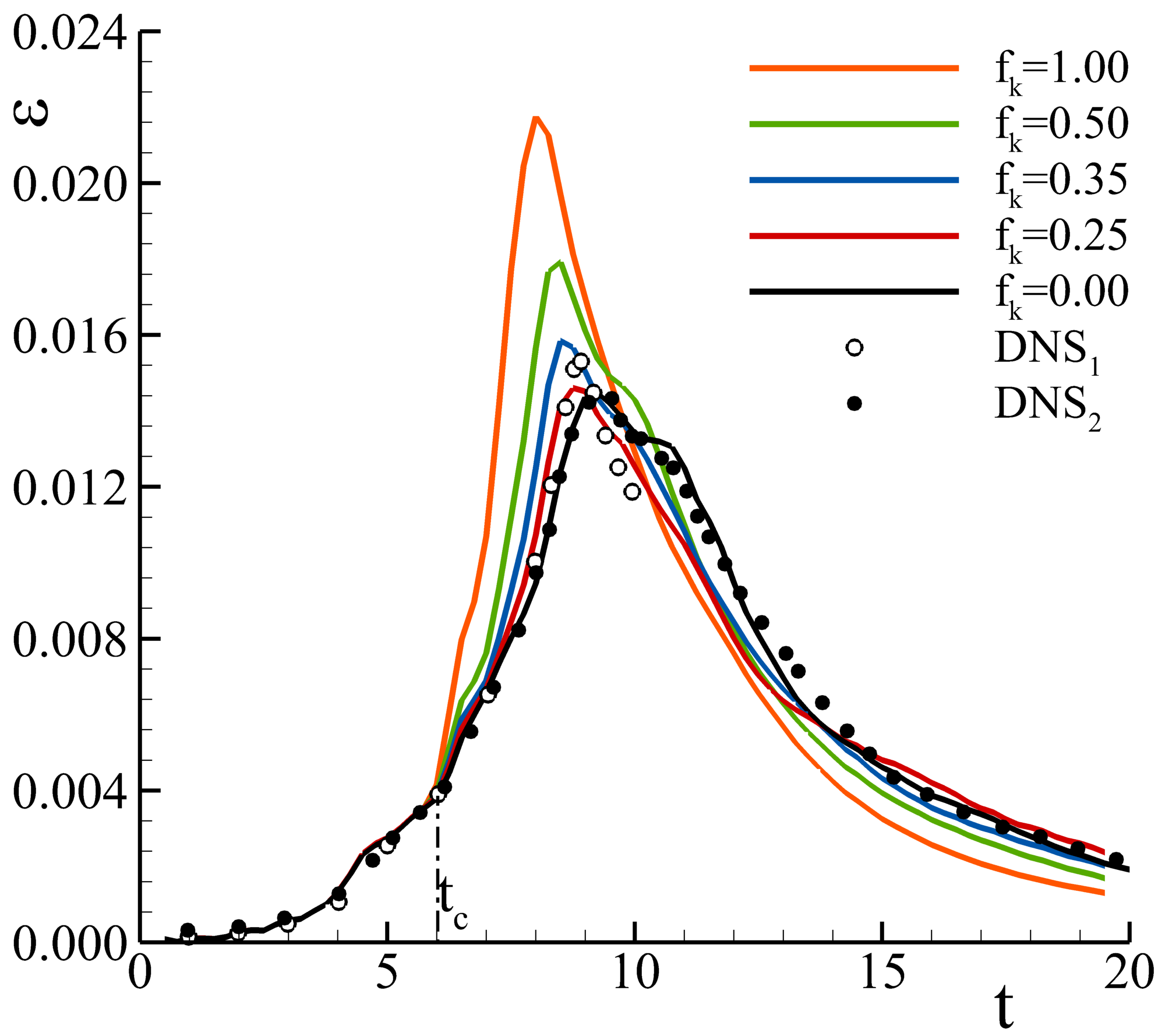}}
\caption{Temporal evolution of the total kinetic energy, $k$, and dissipation, $\varepsilon$, with different $f_k$.  $\mathrm{DNS_1}$, \cite{BRACHET_JFM_1983}; $\mathrm{DNS_2}$, \cite{DRIKAKIS_JOT_2007}.}
\label{fig:4.2_3} 
\end{figure}

Now we turn our attention to figure \ref{fig:4.2_3}, which presents the temporal evolution of $k$ and its dissipation, $\varepsilon$,
\begin{equation}
\label{eq:4.2_1}
\varepsilon = - \frac{d k}{d t}  \; ,
\end{equation}
obtained at different $f_k$ on the finest grid resolution available. The results of $\varepsilon$ are compared to the DNS of Brachet et al. \cite{BRACHET_JFM_1983} ($\mathrm{Ma}=0$, $\mathrm{DNS_1}$) and Drikakis et al. \cite{DRIKAKIS_JOT_2007} ($\mathrm{Ma}_o=0.28$, $\mathrm{DNS_2}$). These reference studies do not provide numerical uncertainties. Figure \ref{fig:4.2_3a} shows that all simulations converge monotonically upon $f_k$ refinement, and those at $f_k \leq 0.35$ exhibit small discrepancies, which do not exceed $0.0047$. The exception in monotonic convergence is observed at $t>11$ at $f_k\leq 0.35$. This result is attributed to numerical uncertainty, especially for the case at $f_k=0.00$ (figure \ref{fig:4.2_2}). The results also show that all computations are independent of $f_k$ until $t\approx t_c= 6$. This is the time when simulations at different $f_k$ lead to non-negligible levels of turbulence stresses.

The results for $\varepsilon$ depicted in figure \ref{fig:4.2_3b} exhibit similar tendencies. Until $t\approx t_c$, all computations are independent of $f_k$ and lead to minor comparison errors. In contrast, the simulations become highly dependent on $f_k$ at $t>t_c$, and their solutions converge upon $f_k$ refinement toward the reference data. Also, solutions of simulations at $f_k \leq 0.35$ are significantly less dependent on $f_k$, and the resulting comparison errors are small. Comparing the two DNS data sets, our results converge toward those of Drikakis et al. \cite{DRIKAKIS_JOT_2007} ($\mathrm{DNS_2}$). This stems from the fact that our simulations match the $\mathrm{Ma}$ of such DNS, highlighting the role of input errors and $\mathrm{Ma}$ \cite{VIRK_JFM_1995} to the comparisons. Regarding simulations at $f_k\ge 0.50$, these lead to large comparison errors. For instance, the magnitude of the peak of dissipation predicted with $f_k=1.00$ is $52.0\%$ larger than that reported by Drikakis et al. \cite{DRIKAKIS_JOT_2007}.

In summary, figures \ref{fig:4.2_1} to \ref{fig:4.2_3} allow us to evaluate the simulations' modeling accuracy through $r_i$ and $f_k$ refinement studies. These converge toward the reference solutions, leading to small comparison errors. The results also reaffirm that the proposed V\&V method can be confidently applied when reference data is unavailable or the flow conditions are unknown since it can generate a reference solution ($r_i\rightarrow 0$ and $f_k\rightarrow 0$). The interpretation of the results also indicates the existence of two groups of simulations: those at $f_k \ge 0.50$ and $f_k<0.50$. This can be explained by the critical value of $f_k$ that allows the SRS model to accurately resolve the key instabilities and coherent structures of the flow, not amenable to modeling. For TGV, these are the vortex reconnection phenomena responsible for the onset of turbulence \cite{PEREIRA_PRF_2021}. 
%
%
%
\subsection{Rayleigh-Taylor}
\label{sec:4.3}
\begin{figure}[t!]
\centering
\subfloat[$h$.]{\label{fig:4.3_1a}
\includegraphics[scale=0.11,trim=0 0 0 0,clip]{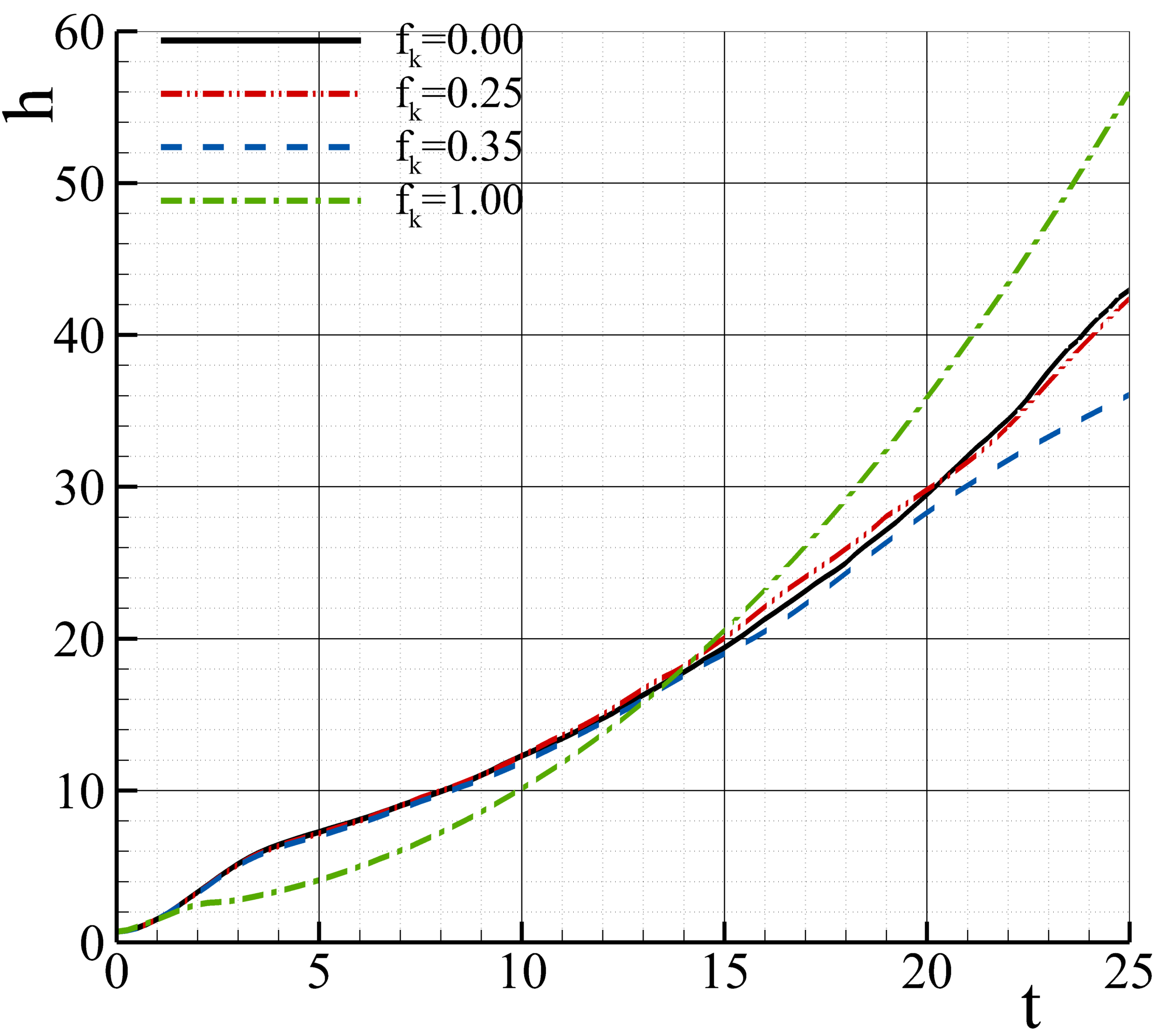}}
~
\subfloat[$\theta$.]{\label{fig:4.3_1b}
\includegraphics[scale=0.11,trim=0 0 0 0,clip]{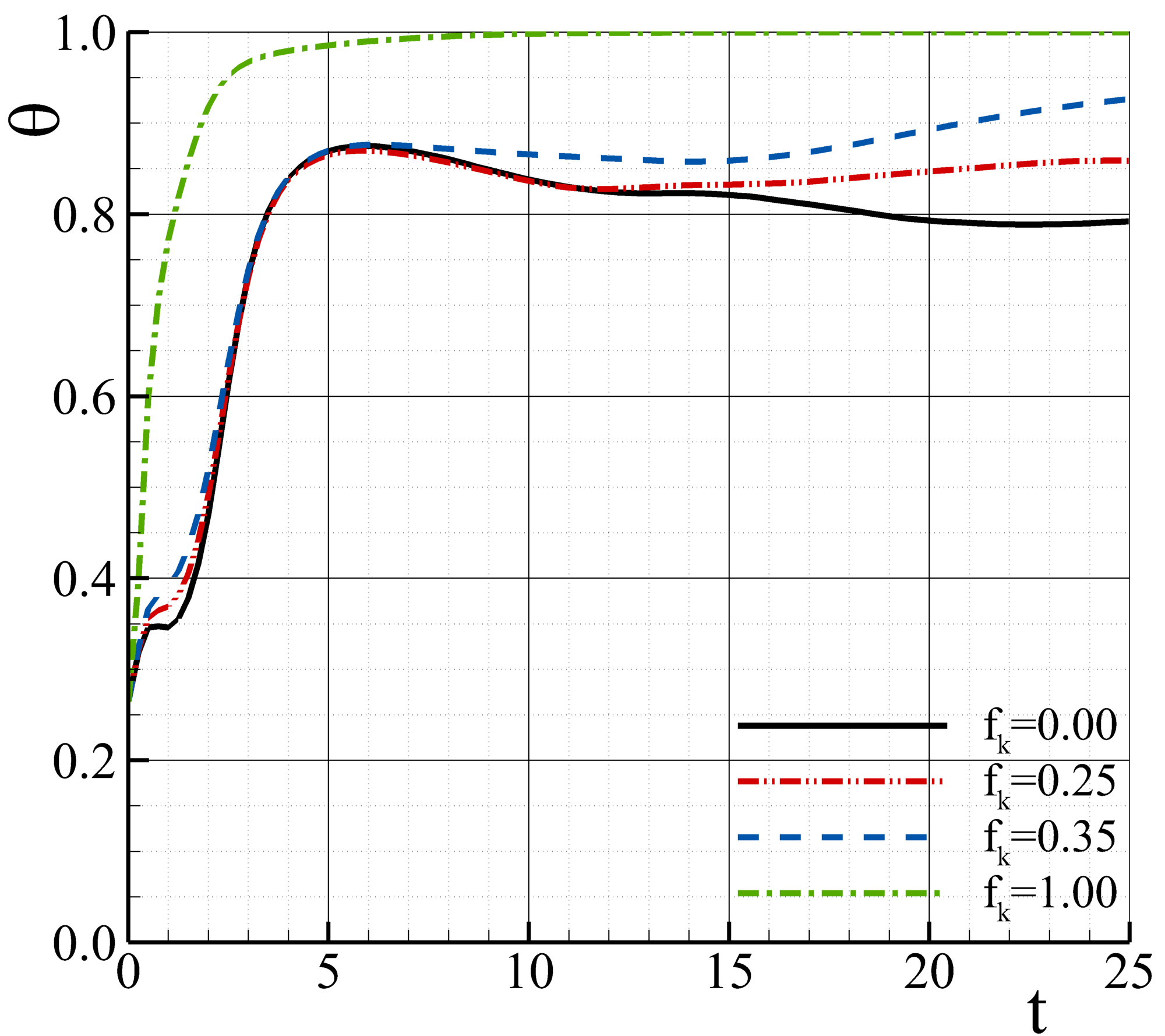}}
\caption{Temporal evolution of the mixing-layer height, $h$, and the molecular mixing parameter, $\theta$, with different $f_k$.}
\label{fig:4.3_1} 
\end{figure}

We conclude this section with a brief presentation of the RT flow, as these results exhibit behaviors similar to those of the CC and TGV. A complete discussion of the current results is given in Pereira et al. \cite{PEREIRA_PRF_2021,PEREIRA_POF_2021}. The RT is a variable-density flow highly dependent on the initial conditions. Hence, we conduct simulations at different values of $f_k$ and compare the results against the case at $f_k=0$ to guarantee the same initial conditions for all values of $f_k$ \cite{DIMONTE_PRE_2004}. Figure \ref{fig:4.3_1} presents the temporal evolution of the mixing-layer height, $h$, and the molecular mixing parameter, $\theta$, predicted with different values of $f_k$ on the finest grid resolution available for all $f_k$ ($N=256^2\times 768$). The second quantity quantifies the homogeneity of the mixing-layer, with $\theta=0$ and $\theta=1$ referring to a heterogeneous and homogeneous mixture, respectively.

The results show that the simulations converge monotonically toward the reference solution upon physical resolution refinement, $f_k \rightarrow 0$. Also, whereas simulations at $f_k \le 0.35$ are in good agreement with the reference solution, those at $f_k \ge 0.50$ lead to large discrepancies. As discussed in Pereira et al. \cite{PEREIRA_PRF2_2021,PEREIRA_POF_2021}, this is caused by the ability of simulations at $f_k < 0.50$ to accurately predict the instabilities and coherent structures governing the flow dynamics. This requires resolving the RT instability, laminar spikes and bubbles, and the Kelvin-Helmholtz instability responsible for the onset of turbulence in the RT flow. Once again, the grid and $f_k$ refinement studies are critical to draw such conclusions.

\begin{figure}[t!]
\centering
\subfloat[$h(t) \pm U_n(t)$.]{\label{fig:4.3_2a}
\includegraphics[scale=0.11,trim=0 0 0 0,clip]{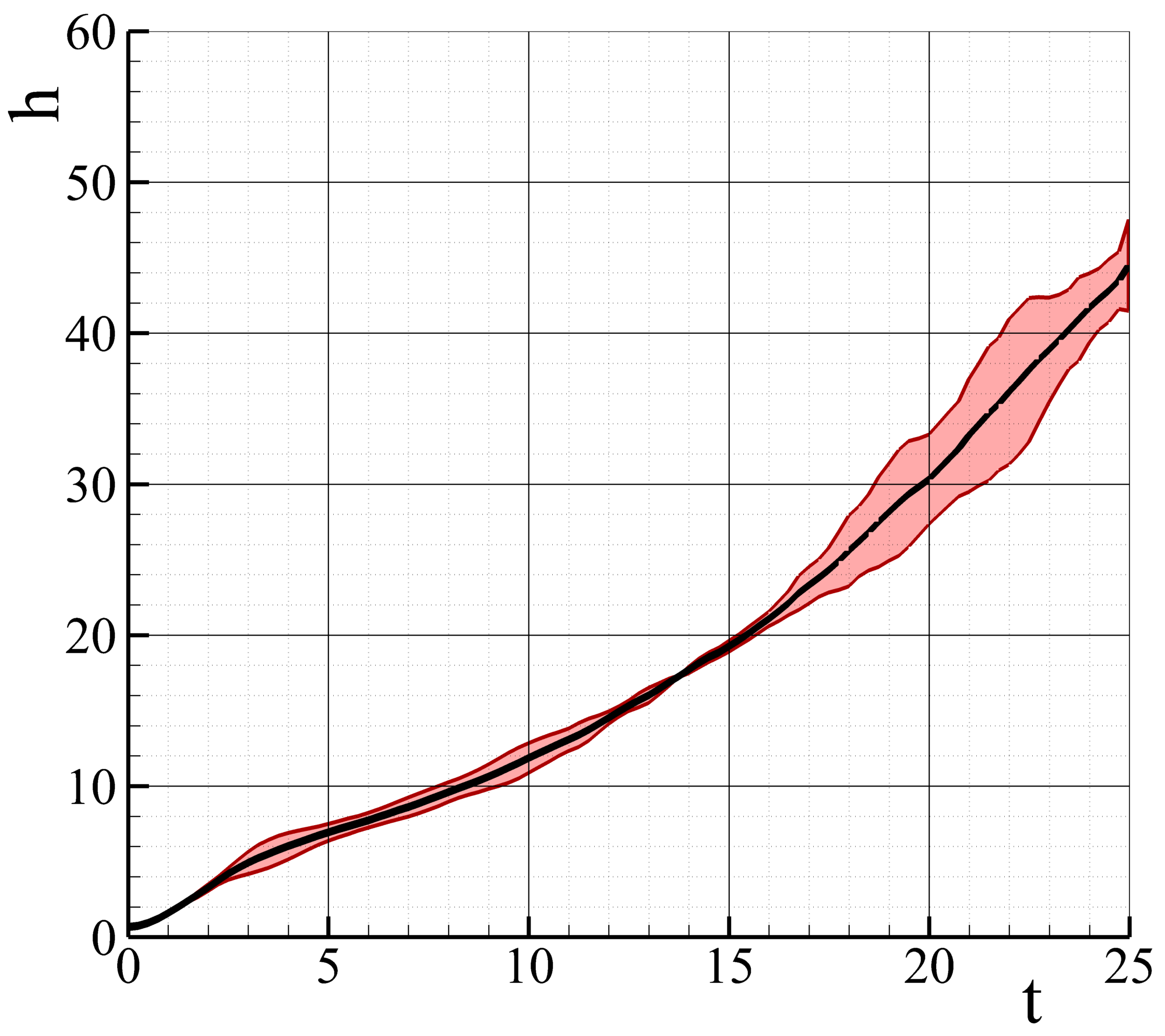}}
~
\subfloat[$h(N)$.]{\label{fig:4.3_2b}
\includegraphics[scale=0.11,trim=0 0 0 0,clip]{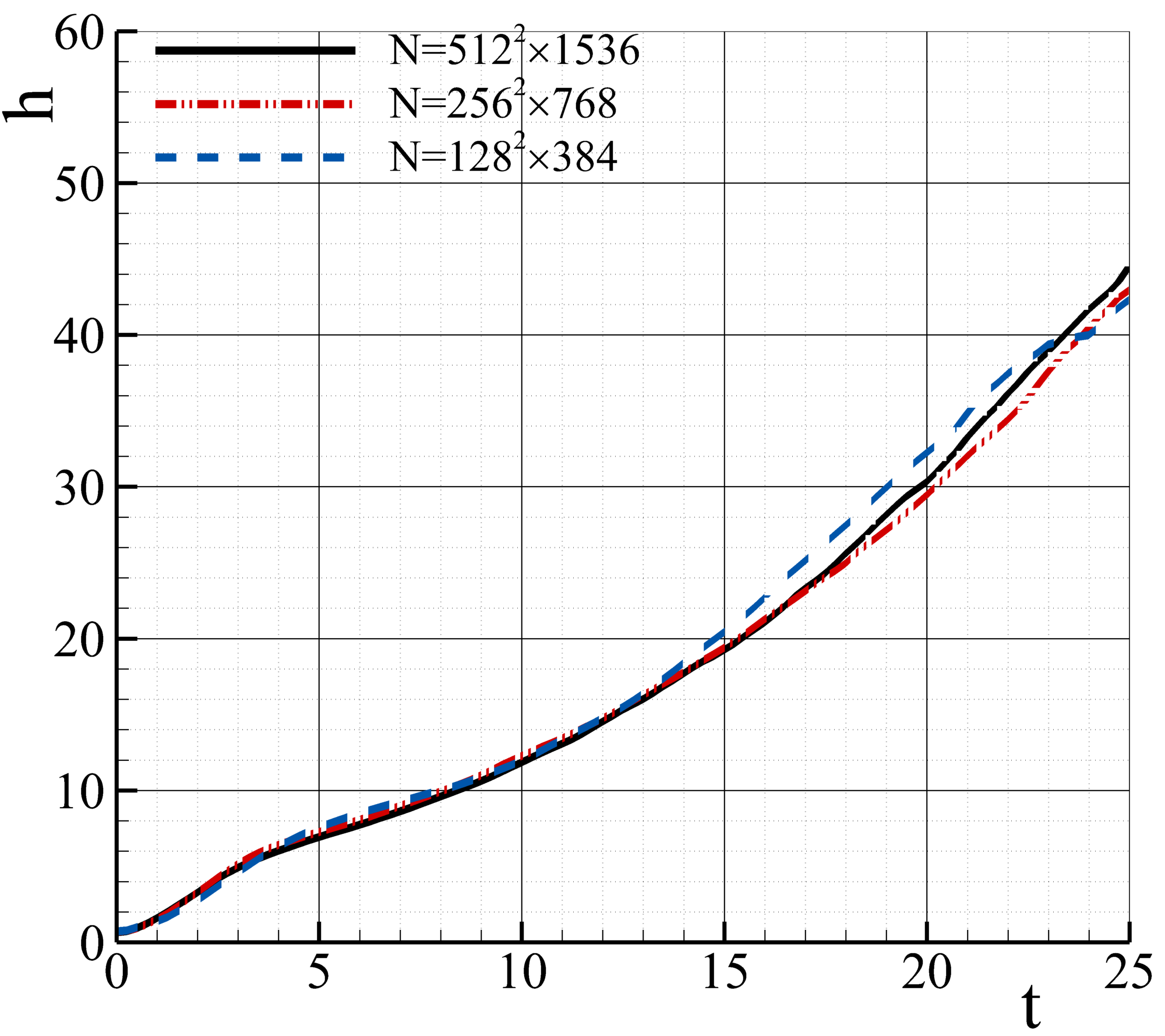}}
\caption{Temporal evolution of the mixing-layer height, $h$, respective numerical uncertainty, $U_n(h)$, and grid resolution dependence for $f_k=0$.}
\label{fig:4.3_2} 
\end{figure}

Finally, figure \ref{fig:4.3_2} presents the numerical uncertainty of $h$ obtained with $f_k=0$ (most demanding case) using the three finest grid resolutions - $N=128^2\times 384$, $256^2\times 768$, and $512^2\times 1536$ cells. The solutions of $h$ for the three grids are also included. The results of figure \ref{fig:4.3_2a} indicate that the numerical uncertainty of the simulations is relatively small, and does not exceed $16.8\%$ of the predicted value during the simulated time. Note that this represents an upper limit for $U_n(h)$ since the coarsest grid used to estimate the numerical uncertainty does not possess sufficient resolution to resolve the interface perturbations at $t=0$. This idea is confirmed in figure \ref{fig:4.3_2b}, which compares the solutions of $h$ for the three grids. Whereas the two finest grids lead to nearly identical solutions, the coarsest one leads to large discrepancies. Such outcome demonstrates the importance of (adequately resolving) the initial flow conditions, and the adequacy of the grid with $256^2\times 768$ cells for the present study.

%
%
%
\section{Conclusions}
\label{sec:5}
This work investigated the importance of V\&V to the credibility and further establishment of SRS of turbulence. A novel but simple V\&V strategy based on grid and physical resolution studies is proposed that permits the separate assessment of numerical and modeling errors. It can be used when reference data is unavailable since it can estimate a reference solution from the physical resolution refinement studies. Also, it permits performing physical resolution refinements using the same initial flow conditions, avoiding the common difficulties generated by under-characterized or unknown experimental conditions. 

To ascertain the former aspects, we simulate three benchmark transitional flows using different grids and physical resolutions: the flow past a circular cylinder at $\mathrm{Re}=3900$, the TGV at $\mathrm{Re}=3000$, and the variable-density RT flow at $\mathrm{At}=0.50$. The results confirm the importance of V\&V to SRS of turbulence, indicate that it is possible to perform reliable V\&V exercises with SRS formulations, and show the potential of the proposed V\&V strategy. It enables an easier identification of cases experiencing error canceling, illustrates that insufficient grid resolution can suppress the modeling advantages of SRS of turbulence, permits a better flow analysis, and the identification of key flow phenomena. These properties can be used to develop better mathematical models and understand the reasons why models fail or succeed.

We expect the proposed V\&V strategy to be a valuable contribution to further develop turbulence modeling, allow users attaining higher-quality computations, and promote the use of V\&V in practical simulations of turbulence. Also, it can be a valuable tool for flows without available reference data or under-characterized initial and flow experimental conditions.  Naturally, performing grid and physical resolution refinement studies is computationally intensive, but the results of this study and figure \ref{fig:1_1} illustrate the risks of not assessing the accuracy of the simulations.
%
%
%
%
\begin{acknowledgment}
We would like to thank the reviewers for their suggestions that improved our paper. Los Alamos National Laboratory (LANL) is operated by TRIAD National Security, LLC for the US DOE NNSA. The research presented in this article was supported by the Laboratory Directed Research and Development program of Los Alamos National Laboratory under project number 20210289ER .
\end{acknowledgment}
%





\end{document}